\documentclass[aps,prd,groupedaddress, twocolumn,  nofootinbib]{revtex4-1}
\usepackage{graphicx, amsmath, bbm, hyperref}
\usepackage{color}

\begin{document}

\title{Hypercuboidal renormalization in spin foam quantum gravity}

\author{Benjamin Bahr}
\email[]{benjamin.bahr@desy.de}

\author{Sebastian Steinhaus}
\email[]{sebastian.steinhaus@desy.de}
\affiliation{II. Institute for Theoretical Physics\\University of Hamburg\\ Luruper Chaussee 149\\22761 Hamburg\\Germany}

\date{\today}

\begin{abstract}
In this article we apply background-independent renormalization group methods to spin foam quantum gravity. It is aimed at extending and elucidating the analysis of a companion letter, in which the existence of a fixed point in the truncated RG flow for the model was reported. Here we repeat the analysis with various modifications, and find that both qualitative and quantitative features of the fixed point are robust in this setting. We also go into details about the various approximation schemes employed in the analysis.
\end{abstract}


\maketitle

\section{Motivation}

The renormalization group (RG) is a pivotal tool in modern physics. It is a method to relate (quantum or statistical) theories defined at different length scales, e.g.~by extracting a macroscopic effective dynamics from a microscopic model. Thus the RG can help in extracting predictions from theories and contrast them with current observations, as well as checking the consistency of the theory \cite{Wilson:1973jj}.

Naturally the RG is of particular interest in quantum gravity, for which usual renormalization techniques run into problems. That is due  to the fact that gravity is not renormalisable as a perturbative quantum field theory \cite{Goroff:1985sz}. This renders the result of the RG flow non-predictive.

With the advent of non-perturbative and background-independent attempts to define a theory of quantum gravity (see \cite{Oriti:2009zz} for an overview to several different approaches), the question arises how these models behave under renormalization. In this article we pursue this question, and focus on so-called spin foam models, which are a path integral approach related to loop quantum gravity, and thus sometimes referred to as covariant loop quantum gravity in the literature \cite{thomasbook, cbook, Perez:2012wv}.

Spin foam models are defined on a discretisation of the space-time manifold. These are usually given by a triangulation \cite{Barrett:1997gw, Engle:2007wy, Freidel:2007py}, although there are extensions which work for a general 2-complex \cite{Kaminski:2009fm}. Said complex itself only carries combinatorial information, and can be regarded as an irregular lattice. Geometric information about the lattice is encoded by coloring it with group theoretic data. Eventually, the colorings are being summed over, which is interpreted as path integral over (discrete, quantum) geometries of spacetime. Thus spin foam models are similar to lattice gauge theories \cite{Bahr:2010bs, Dittrich:2012er}. However, while the latter are explicitly defined on a lattice with a notion of length scale, in spin foam models an a priori numerical scale is absent. At first sight this might obstruct the application of a renormalization scheme, as the scale is usually employed to hierarchically order the (infinitely many) degrees of freedom of the system. However, progress in recent years has outlined a promising route in the context of spin foam models \cite{Oeckl:2002ia, Manrique:2005nn, Rovelli:2010qx, Perini:2008pd, Riello:2013bzw, Dittrich:2012jq, Bahr:2014qza, Dittrich:2014ala, Banburski:2014cwa}.

Before we go into the details of the system studied in this article, we would briefly like to stress the importance and potential lying in the RG for spin foam models:
\begin{itemize}
\item The introduced discretisation plays the role of a fiducial regulator truncating the number of degrees of freedom in a spin foam. Generically the results will depend on the chosen discretisation. As a tool relating models on different discretisations, the RG is therefore essential in checking the consistency of the model.
\item The breaking of diffeomorphism symmetry in discrete models \cite{Dittrich:2008pw} is closely entangled with discretisation (in)dependence, as detailed in \cite{Bahr:2009ku, Bahr:2009qc}. In particular, it has been suggested that it can be restored at RG fixed points \cite{Bahr:2011uj}. Indeed as the fundamental symmetry of general relativity, its fate (at least at an approximate level) is a pivotal question in discrete quantum gravity.
\item In order to extract predictions from a theory it must be possible to efficiently and reliably calculate physical processes. For this it is beneficial to work with discretizations using not too many building blocks. Understanding the RG flow and the behaviour of the theory for lattices of different sizes is an important step to support results which have already been achieved on small triangulations, e.g.~for the computation of cosmological expansion or black hole evaporation \cite{Bianchi:2010zs,Rennert:2013pfa,Christodoulou:2016vny}.
\end{itemize}

It should also be noted that the methods presented here are not the only approaches to renormalizing a quantum theory of general relativity. There are several which make use of slight alterations of the usual quantum field theory methods, such as the effective action approach (see e.g.~\cite{Donoghue:2012zc}), and most notably, the asymptotic safety approach (AS, see \cite{Niedermaier:2006wt} for an overview). It is in particular the latter which postulates the existence of a non-Gaussian fixed point of the RG flow. We discuss connections to AS later in the article.

We would like to stress that in this article we follow the point of view of renormalizing spin foams via refining / coarse graining, i.e. relating models defined on discretisation, which is more akin to the notion of renormalization in lattice gauge theories. However it is still an open issue in the literature whether spin foams should be refined or summed over, which leads to very different assumptions and posed questions. The most holistic approach to sum over spin foams is 
the group field theory approach (GFT, \cite{Oriti:2013aqa}), which takes the interpretation of the spin foam amplitudes as Feynman graphs of the theory literally, and builds its renormalization theory starting from there. The theory is therefore cast into a generalization of matrix models called a tensor field theory \cite{Rivasseau:2011hm}. Statements about renormalization in this framework are subject of extensive current research \cite{Carrozza:2016vsq, Geloun:2016qyb}.

\subsection{Background-independence and renormalization}

Generically in models based on a discretisation, the RG relates theories with different numbers of degrees of freedom, e.g. under coarse graining. Crucially both theories describe the same physical situation, where the coarser can be seen as an effective theory, where some of the finer degrees of freedom have been integrated out. An absolute scale nicely encapsulates this hierarchy, but in fact to study a RG flow, just a relative scale is sufficient. To do so we shift the perspective away from an absolute scale towards relating different discretisations to one another, as well as theories defined on them.

In spin foam models the physical information is stored in transition amplitudes between states encoding 3D geometries on the boundary of space-time, which usually play the role of Cauchy surfaces. The discretization of the bulk induces one of the boundary. Since these discretizations can be successively refined, the same (discrete) 3D geometry can be represented differently, while still representing the same state. Therefore, one needs to be able to relate spin foam models on different(ly discretized) boundaries, such that they give the same transition amplitude. This defines the RG flow of the model, i.e.~from finer to coarser discretizations.

States defined on different boundaries are related to one another by embedding maps. These embed a state represented on a coarse boundary and thus `coarser' Hilbert space, into a finer boundary and thus `finer' Hilbert space. For this to be well-defined, such that states can be unambiguously related, the embedding maps have to satisfy several consistency conditions. In an inductive limit construction one obtains a continuum Hilbert space, which in turn allows for a definition of a continuum limit of spin foam models. Successful constructions of such embedding maps and families of Hilbert spaces in quantum gravity are the Ashtekar-Lewandowski representation \cite{Ashtekar:1991kc, Ashtekar:1994mh} and the recently developed BF representation \cite{Dittrich:2014wpa, Bahr:2015bra, Dittrich:2016typ}.

In this RG procedure the choice of embedding maps plays a key role as they define how degrees of freedom are coarse grained  \cite{Dittrich:2011ke, Dittrich:2013jaa, Hoehn:2014fka, Hoehn:2014wwa}. Therefore it was argued in \cite{Dittrich:2013xwa} that these embedding maps should be compatible with the dynamics of the theory to be studied, that is the dynamics of the theory should determine how states are refined, which one can regard as a dynamical vacuum, and how states should be coarse grained.

This scheme is precisely realized in tensor network renormalization: there one expresses the partition function of the system as a contraction of a tensor network, i.e. a network of multi-dimensional arrays \cite{guwen, levin, vidal-evenbly}. These tensors are locally manipulated to obtain a coarser network which approximates the original partition function as well as possible. Thus this scheme describes a RG flow of tensors. In order to derive the new tensors one performs variable transformations from fine to coarse tensor indices, i.e. the inverse of an embedding map, where the coarse indices are ordered by their significance to the dynamics, such that a suitable truncation is possible. This RG scheme was applied to so-called spin net models, 2D models analogue to spin foams, and uncovered a rich phase structure \cite{Dittrich:2011zh, Dittrich:2013bza, Dittrich:2013voa}. Most recently it was applied to models constructed analogous to $4D$ spin foam models, which revealed a complex and intriguing dynamics achieved by the imposition of simplicity constraints \cite{Dittrich:2016tys}. Also the algorithm has been generalized to also tackle lattice gauge theories and spin foam models \cite{Dittrich:2014mxa,Delcamp:2016dqo}.

\subsection{Cuboidal spin foams}

Despite these encouraging recent developments, renormalizing $4D$ spin foam models remains a challenge, predominantly because of their algebraic complexity. As soon as one considers spin foams consisting of several building blocks without further approximations or simplifications, numerical or analytical techniques alone are not efficient to extract new insights about the theory. A combination of both methods (including suitable approximations) seems to be a promising route towards this goal. The new representation of loop quantum gravity, called the BF vacuum \cite{Dittrich:2014wpa, Bahr:2015bra, Dittrich:2016typ}, and the `fusion basis' \cite{Delcamp:2016yix} appear as interesting first steps towards better understanding the theory from the analytical perspective. In this article we take a different approach, where we consider a drastically restricted version of the full $4D$ theory, which notably are part of the full $4D$ spin foam path integral, and use the geometry in order to relate finer and coarser spin foams.

We study the Engle-Pereira-Rovelli-Livine / Freidel-Kransov (EPRL-FK) model defined on a hypercubic two-complex \cite{Engle:2007wy, Kaminski:2009fm}. Instead of studying it in full generality, we constrain the 2-complex to be (hyper)cuboid-shaped. More concretely this entails a restriction on the allowed geometric data that are summed over in the path integral, to allow only (hyper-)cuboidal quantum geometries, which have been studied in \cite{Bahr:2015gxa}. As the analysis performed in this article is based on this previous work, let us briefly recall its main results.

In \cite{Bahr:2015gxa} we defined the amplitudes of the (hyper-)cuboidal spin foam model and computed its asymptotic expansion, which we explicitly give in appendix \ref{Sec:Appx:Amplitude}. We examined several qualitative features of this amplitude, most notably the fate of diffeomorphism symmetry. To this end we considered two glued hypercuboids at fixed total volume and studied how the amplitude changes under shifting the 3D hypersurface along which they are glued. We interpreted these vertex translations as (an Abelian subgroup of discrete) diffeomorphisms similar to \cite{Bahr:2011uj}. Generically the model is not invariant under these shifts, yet the parameters of the model can be chosen such that this symmetry is (almost) restored. We recall this analysis in section \ref{Sec:Results:VTS}.

In the present article we build on the results of \cite{Bahr:2015gxa} and use the same setup to extract {\it dynamical} features of the model. Via Monte-Carlo methods we compute the (variance of the) volume of a hypercuboid as an observable for a coarse and a refined spin foam describing the same geometry. From this observable we define the renormalization group flow of spin foam amplitudes from fine to coarser foams and find indications for a UV-attractive fixed point.

The restriction to hypercubic geometries comes at a price, as the hypercuboids are flatly glued together by the dynamics; thus these spin foams cannot capture curvature. Indeed in the asymptotic expansion thoroughly studied in \cite{Bahr:2015gxa}, the Regge action associated to the hypercuboid (including the boundary terms) always vanishes.

The fact that cuboidal spin foams cannot describe curvature degrees of freedom is a very drastic approximation, which is not expected to be valid in situations where high curvature plays an important role. Nevertheless it is interesting to study them in more detail for various reasons.
\begin{itemize}
\item Quantum cuboids are no mere toy model as they are actual configurations occurring in the full $4D$ spin foam path integral. Moreover in particular physical situations, e.g. close to flat space-times, they might significantly contribute to the state sum\footnote{A priori it is not clear whether the path integral is dominated by configurations with highest weight, as they might only represent a measure zero set.}. Indeed, the RG methods described in this article provide a first step towards checking this claim.
\item Due to the geometric construction with coherent intertwiners, the model can be generalized straightforwardly `around' cuboidal states. Such deviations from cuboids will allow for curvature degrees of freedom, while still containing the cuboid case. Thus the results presented in this article can be checked for consistency.
\item The simplicity of quantum cuboids, both in their geometric interpretation and algebraic complexity, allows for a first implementation of the RG methods outlined in \cite{Bahr:2014qza, Dittrich:2014ala, Bahr:2015tne}. Thus we can develop and test numerical implementations of these schemes, which are a good foundation to expand upon.
\end{itemize}

In addition to these arguments, first results of renormalizing quantum cuboids are very promising. In \cite{Bahr:2016hwc} we related coarse and fine hypercuboids geometrically; essentially a collection of fine hypercuboids should `behave' like a coarse hypercuboid with the areas of fine rectangles summing up to the area of a coarse rectangle. We studied said integral with Monte Carlo techniques, and computed expectation values of observables. By comparing these observables for the original and renormalized amplitudes, we derived a renormalization group flow by essentially projecting the renormalized amplitude back to the original one for a different parameter of the model. This projected RG flow revealed first indications of a UV fixed point, separating two qualitatively different phases, one in which very irregular subdivisions dominated and one in which regular subdivisions gave the largest contribution. Right on the fixed point almost all configurations contribute the same, however, which is a strong indication that diffeomorphism symmetry might be restored on this point.

In this article we will more thoroughly discuss the RG scheme and confirm previous results. In addition to that we also present properties of the renormalized amplitude itself. This is possible thanks to numerical algorithms suited to multidimensional integrals \cite{Hahn:2004fe}, some of which are also straightforwardly applicable to highly oscillating systems. Therefore we expect them to be successfully applicable also in the case of non-vanishing curvature.

This paper is organized as follows: In section \ref{Sec:Model}, we will recap the definition and properties of the Riemannien signature EPRL-FK model, which we will work with for the rest of the article. In section \ref{Sec:Renzn}, we will discuss the background-independent renormalization approach which is used in spin foam models. We have reserved all of section \ref{Sec:Appr} to discuss which kind of approximations we are going to employ, in order to compute the renormalization group flow. In section \ref{Sec:Results} we will then go over the results of our numerical analysis of the model, which we obtained using the approximations discussed before. We conclude the main part of the article with a summary and discussion section\ref{Sec:SummAndDiss}. In the appendix, we go deeper into technical details of the model itself, and provide some explicit calculations behind our analysis, as well as a brief comment about the employed numerical methods.

\section{Spin Foam Models}\label{Sec:Model}
\subsection{The model}

\begin{figure}[hbt!]
\includegraphics[width=0.30\textwidth]{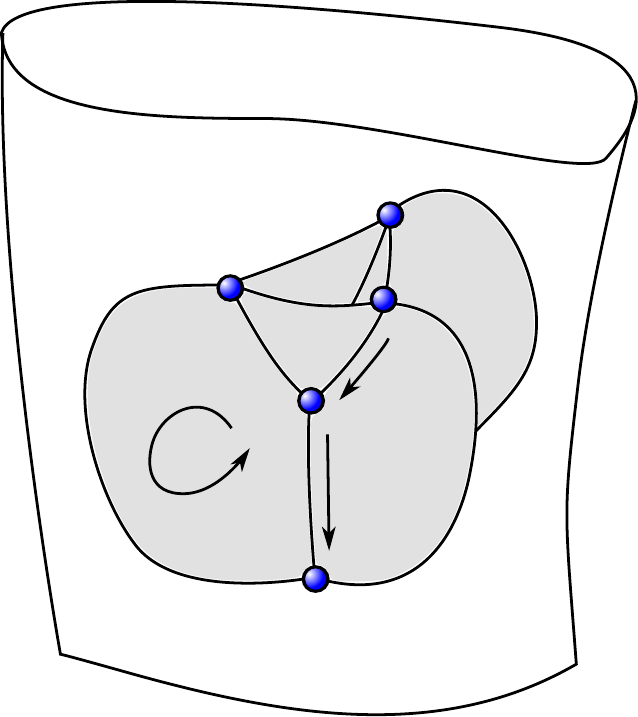}
\caption{A $2$-complex, embedded in the space-time manifold, is defined by vertices, edges and faces, glued together according to specific rules. The latter two are supposed to carry an orientation, on which the resulting spin foam amplitude does not depend, however. Note there is no metric on the manifold, so there is no way of saying ``how long'' a specific edge is, for instance.}\label{Fig:figure_2complex}
\end{figure}

In this article we consider the Riemannian signature EPRL-FK spin foam model with Barbero-Immirzi parameter $\gamma < 1$. The EPRL-FK model describes a dynamics evolving spin network states of loop quantum gravity situated on the boundary of the foam. These states are represented on graphs $\Gamma$, while the spin foam is defined on a 2-complex mediating between the boundary graphs. The 2-complex is prescribed entirely by its vertices $v$, edges $e$ and faces $f$ (and their combinatorics) and by itself does not carry any geometric interpretation, in particular no background geometry (see figure \ref{Fig:figure_2complex})

\begin{figure}[hbt!]
\includegraphics[width=0.30\textwidth]{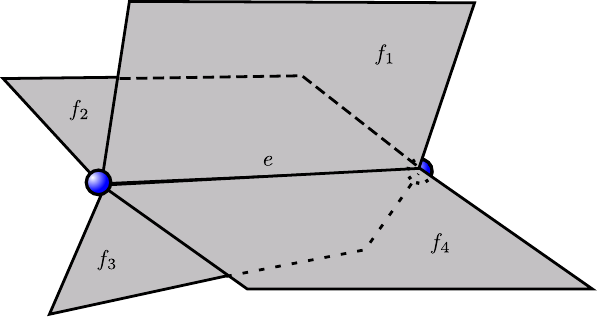}
\caption{Intertwiners $\iota_e$ are vectors in the ege Hilbert spaces $\iota_e\in\mathcal{H}_e$. Assuming that all orientations of the $f_i$ agree with the one on $e$, the edge Hilbert space is given by $\mathcal{H}_e=\text{Inv}_{SU(2)}(V_{j_1}\otimes\cdots V_{j_n})$. For each face in which the orientation does not agree with the one on $e$, the representation space $V_{j_f}$ is replaced by its dual.}\label{Fig:EdgeHilbertSpace}
\end{figure}

A (pre)geometric interpretation is endowed onto the foam by colouring it with group theoretic data, irreducible representations $j_f\in\frac{1}{2}\mathbbm{N}$ of the gauge group $SU(2)$ to the faces $f$ and intertwiners $\iota_e$ to the edges $e$, i.e. vectors in the invariant subspace of the tensor product of representation spaces assigned to faces meeting at that edge (see figure \ref{Fig:EdgeHilbertSpace}). Such an assignment $\{j_f,\iota_e\}_{f, e}$ is called a spin foam state. Geometrically the intertwiners are interpreted as the dual of 3D fuzzy polyhedra, where the areas of its faces are given by the representations meeting at the edge. The 2-complex then combinatorially describes how these 3D building blocks are glued to form a $4D$ geometry. Thus the spin foam state is a possible geometry interpolating between the 3D geometries encoded in the boundary spin network state.

As a path integral approach, the physical content of spin foam models is encoded in its transition amplitudes. To this end an amplitude is assigned to each spin foam state and then one sums over all possible spin foam states. In current spin foam models this assignment is local, by assigning a vertex amplitude $\mathcal{A}_v$ to vertices $v$, an edge amplitude $\mathcal{A}_e$ to edges $e$ and a face amplitude $\mathcal{A}_f$ to faces $f$. The partition function of the spin foam is given by:
\begin{equation}\label{Eq:SF_StateSum}
Z_\Gamma = \sum_{j_f, \iota_e} \prod_v \mathcal{A}_v \prod_e \mathcal{A}_e \prod_f \mathcal{A}_f
\end{equation}

The model depends on the Barbero-Immirzi parameter $\gamma\in\mathbbm{R}\backslash\{0,\pm 1\}$, which needs to be such that
\begin{eqnarray}
j^\pm_f:=\frac{1}{2}|1\pm\gamma|j_f
\end{eqnarray}

\noindent are half-integers as well.\footnote{If there are no half-integer solutions for $j^\pm_f$, then the amplitude is defined to be zero, so effectively, $j_f$ does not appear in the sum (\ref{Eq:SF_StateSum}). In particular, if $\gamma$ is irrational, the model is trivially $Z=1$. Also, depending on the fractional representation of a rational $\gamma$, quite a few spins might be excluded in the sum. This is a specific feature of the Euclidean signature model, which does not appear for Lorentzian signature \cite{Pereira:2007nh}. Since, for most of the article, we are interested in the large $j$-limit, where the $j_f$ essentially become continuous, we can ignore this point from now on.}

\subsection{Amplitude functions}

The main ingredient of the model are the amplitude functions $\mathcal{A}_f$, $\mathcal{A}_e$, and $\mathcal{A}_v$. To define these, we need the so-called EPRL maps $Y_e^\gamma$, which are defined for each edge $e$, depending on the spins $j_1,\ldots j_n$ attached to all faces $f$ touching $e$.
\begin{eqnarray}
Y_e^\gamma\;&:&\;\text{Inv}_{SU(2)}(V_{j_1}\otimes \cdots \otimes V_{j_n})\;\\[5pt]\nonumber
&&\qquad\qquad\to\;\text{Inv}_{\text{Spin}(4)}(V_{j^+_1,j_1^-}\otimes \cdots \otimes V_{j_n^+,j_n^-}),
\end{eqnarray}

\noindent where the isomorphism $\text{Spin}(4)\simeq SU(2)\times SU(2)$ is used. The map is defined as follows: for each half-integer $j$, define
\begin{eqnarray}
\beta_j^\gamma:V_j\;\to\;V_{j^+,j^-}
\end{eqnarray}
\noindent by the unique isometric embedding of $V_j$ into the factor which appears in the Clebsch-Gordan-decomposition of $V_{j^+,j^-}\simeq V_{j^+}\otimes V_{j^-}$. Denote by $P$ the projection onto the $\text{Spin}(4)$-gauge-invariant subspace of $V_{j^+_1,j_1^-}\otimes \cdots \otimes V_{j_n^+,j_n^-}$, then
\begin{eqnarray}
Y_e^\gamma\;:=\;P\Big(\beta_{j_1}^\gamma\otimes\cdots \otimes \beta_{j_n}^\gamma\Big).
\end{eqnarray}

\noindent The amplitude functions are defined as follows: The face amplitude $\mathcal{A}_f$ depends on $j_f$ and a coupling constant $\alpha\in\mathbbm{R}$, and is given by
\begin{eqnarray}
\mathcal{A}_f\;:=\;\Big((2j^++1)(2j^-+1)\Big)^\alpha.
\end{eqnarray}

\noindent The edge amplitude is chosen to be simply the normalization of the intertwiners, i.e.
\begin{eqnarray}
\mathcal{A}_e\;:=\;\frac{1}{\|Y_e^\gamma\iota_e\|^2}.
\end{eqnarray}

\begin{figure}[hbt!]
\includegraphics[width=0.30\textwidth]{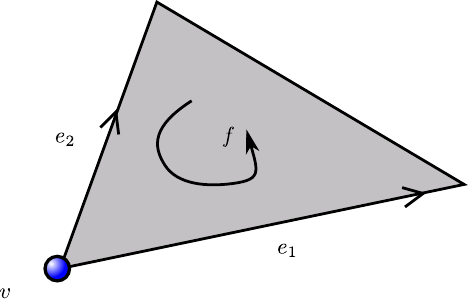}
\caption{The edge Hilbert spaces $\mathcal{H}_{e_{1,2}}$ are such that $\iota_1$ and $\iota_2$ both have have one index which belongs to the representation $j^\pm_f$ at the face $f$. Irrespective of the orientation, these indices are always in opposite positions, so they can be unambiguously contracted, to form the vertex amplitude $\mathcal{A}_v$. }\label{Fig:VertexContraction}
\end{figure}

\noindent The most involved is the vertex amplitude, which is given by
\begin{eqnarray}
\mathcal{A}_v\;:=\;\text{tr}_v\Big(\bigotimes_{e\supset v}(Y_e^\gamma\iota_e)\Big),
\end{eqnarray}

\noindent where the \emph{vertex trace} $\text{tr}_v$ is defined as follows: In the tensor product of all $Y_e^\gamma\iota_e$, each face $f$ which ends at the vertex $v$ contributes two factors, one each from the two edges which border $f$ at $v$ (see figure \ref{Fig:VertexContraction}). These two ``legs'' of the tensor are contracted with the $\epsilon$-tensor in the $V_{j_f^+}\otimes V_{j_f^-}$-representation. The combinatorics at each vertex is such that the result is a number, which constitutes the vertex amplitude.

There are several generalization of this spin foam model. Apart from working with Lorentzian signature \cite{Pereira:2007nh}, one can include a cosmological constant, e.g.~by replacing the local gauge groups with its quantum group counterparts \cite{Smolin:1995vq, Major:1995yz, Fairbairn:2010cp, Han:2011aa,Dupuis:2013haa, Haggard:2014xoa}. In the future, it might be also interesting to extend the model by including new coupling parameters for terms modelling higher powers of $R_{\mu \nu}$. These terms might give rise to Regge discretisations of higher curvature terms in the large $j$-limit. This is in particular attractive in lieu of the renormalization group flow of the model.

\subsection{The face amplitude and $\alpha$}

A remark on the coupling constant $\alpha$ is in order: in the original EPRL model, the edge- and face amplitudes were not explicitly specified. We want to keep this point open, and have more freedom in our model, which is why we include the parameter $\alpha$. Investigations of how spin foam models  behave depending on $\alpha$ have occurred before in the literature, e.g.~\cite{Baez:2002aw}, \cite{Livine:2006it}, or \cite{Bonzom:2013ofa}, in particular with regards to the convergence of the model.

In the later analysis of this model, we will find that the properties of the model, indeed, crucially depend on the value of $\alpha$.

\section{Renormalization}\label{Sec:Renzn}
\subsection{The notion of scale}\label{Sec:Renzn:Scale}
Renormalization is a central part in quantizations of theories with infinitely many degrees of freedom. In the Wilsonian sense, it is concerned with ordering the degrees of freedom along a certain hierarchy (the scale), and determining the effective dynamics at each scale. This is realized by computing the flow of coupling constants, i.e.~parameters which specify the action governing the effective degrees of freedom. These coupling constants therefore become scale-dependent.  Computing the renormaliziation of a theory therefore amounts to computing the change of coupling constants as the scale changes, i.e.~the ``flow'' of the effective dynamics.

\begin{figure}[hbt!]
\includegraphics[width=0.45\textwidth]{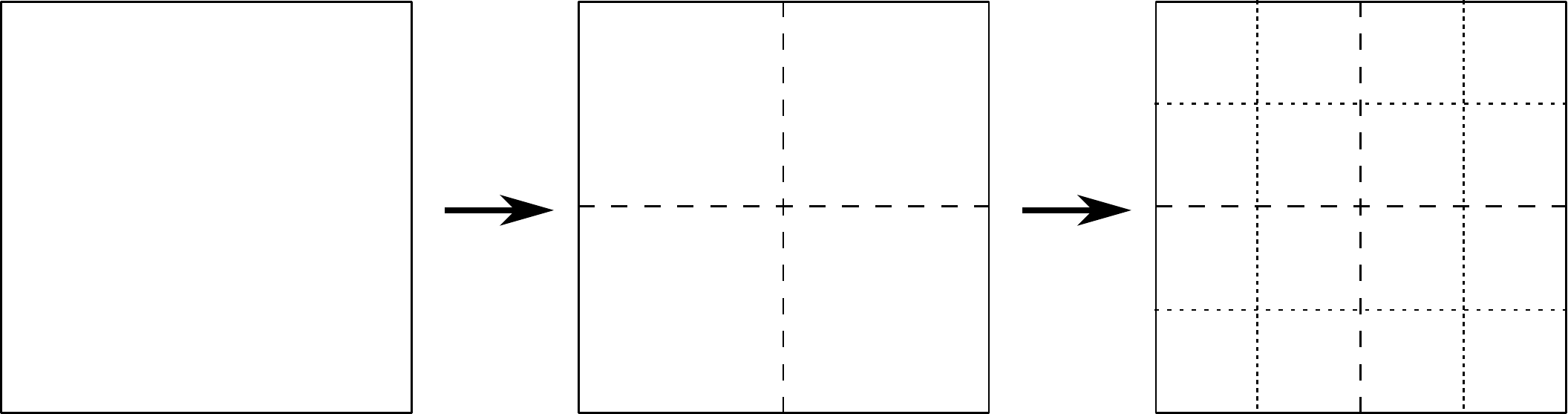}
\caption{Nested lattices, each of which arises as a subdivision of the former one into $2^d$ times as many blocks. In our case, $d=4$.}\label{Fig:NestedLattices}
\end{figure}

In background-dependent theories, the scale is just given by a number (usually a lattice length or momentum-cutoff), so the RG flow generates trajectories through ``theory space''. In the background-independent context of spin foam models, however, one way of defining a scale is by the 2-complex itself \cite{Bahr:2014qza, Bahr:2015tne}\footnote{In the Causal Dynamical Triangulations (CDT) approach, which also does not have a background structure, there is a length scale present in the model, since in the path integral all edge lengths of simplices are being kept fixed, while the sum ranges over different lattices \cite{Ambjorn:2014gsa}.}. Since these do not form a linear, but rather a partially and directed hierarchy, the RG flow runs along a filter, rather than a linear sequence. However, if one restricts to only a subset of 2-complexes, as we do in this work, it is possible to select a subsequence of the set of all 2-complexes. This is, in our case, a sequence of $4D$ hypercubic lattices, which are nested inside each other (see figure \ref{Fig:NestedLattices}). Note that these lattices still do not carry a numerical scale, i.e.~there is no a priori lattice spacing. So it does not make sense to say that, for instance, an edge $e$ is split ``in the middle''. Rather, since geometric information on the lattice is included in the degrees of freedom themselves, the ``length scales'' are related by embedding maps.

\subsection{Embedding maps}\label{Sec:Renzn:EmbMaps}

A crucial part of renormalization, which becomes singularly important in the background-independent context, is the relation of degrees of freedom on different lattices. This is connected to the ``rescaling'' of degrees of freedom in the traditional context, and to the ``block spin transformations'' in the lattice theory context.

One way of describing path integral theories is by the definition of their boundary amplitudes. The boundary states are contained in the associated Hilbert space $\mathcal{H}$, and the amplitude function $\mathcal{A}$ is a linear functional on (possibly a dense subset of) $\mathcal{H}$, assigning transition amplitudes $\mathcal{A}(\psi)$ to states $\psi\in\mathcal{H}$.

\begin{figure}[hbt!]
\includegraphics[width=0.45\textwidth]{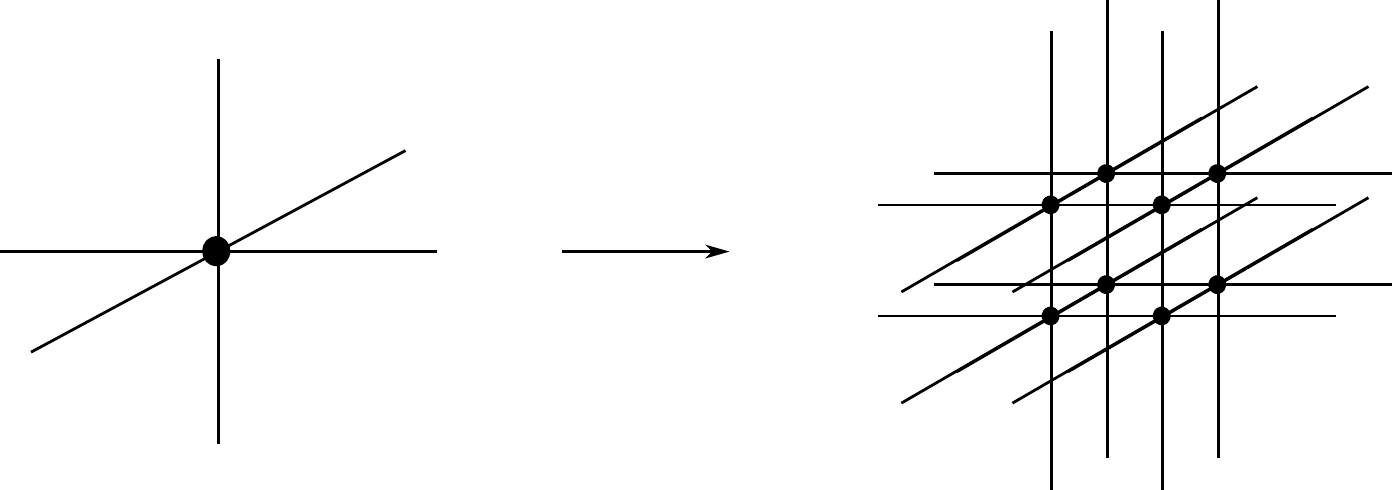}
\caption{A node of a coarse boundary graph $\Gamma'$ is embedded into a collection of eight nodes, which are part of a finer boundary graph $\Gamma$.}\label{Fig:RefiningBoundary}
\end{figure}

In spin foam models, the boundaries of $2$-complexes are given by their respective boundary graphs $\Gamma$, to which the respective boundary Hilbert space $\mathcal{H}_\Gamma$ contains all spin network states on $\Gamma$. The spin network functions on different lattices are related to one another, by the so-called embedding maps. If $\Gamma$ is a boundary graph and $\Gamma'$ a coarser boundary graph in the sense of figure \ref{Fig:RefiningBoundary}, then we write $\Gamma'\leq\Gamma$. The associated embedding map is an isometry
\begin{eqnarray}\label{Eq:EmbeddingMaps}
\iota_{\Gamma\Gamma'}\;:\;\mathcal{H}_{\Gamma'}\;\longrightarrow\;\mathcal{H}_{\Gamma}
\end{eqnarray}

\noindent satisfying $\iota_{\Gamma\Gamma'}\iota_{\Gamma'\Gamma''}=\iota_{\Gamma\Gamma''}$. A cylindrically consistent theory is described in terms of amplitudes $\{\mathcal{A}_\Gamma\}_\Gamma$, satisfying
\begin{eqnarray}\label{Eq:RenormalizedAmplitude_abstract}
\mathcal{A}_{\Gamma'}\;=\;\mathcal{A}_{\Gamma}\iota_{\Gamma\Gamma'}.
\end{eqnarray}

\noindent Traditionally, a theory is called ``renormalizable'' if one can find a parameterization of amplitudes $\mathcal{A}_\Gamma$ in terms of finitely many coupling constants $g_a$, such that, given $g_a$, (\ref{Eq:RenormalizedAmplitude_abstract}) can be realized by an adjustment $g_a\mapsto  g'_a$. One also calls these actions ``form-invariant''. Whether this can be realized in quantum gravity is one of the interesting open questions in the field.

\section{Approximations}\label{Sec:Appr}

In general, it will be nigh impossible to solve the RG flow, i.e.~find $\mathcal{A}_\Gamma$ for all $\Gamma$ such that (\ref{Eq:RenormalizedAmplitude_abstract}) is satisfied. Therefore, we resort to approximations. We will use several major approximations in this article, which we briefly list in what follows.
\begin{itemize}
\item Firstly, we do not consider all possible 2-complexes, but restrict ourselves to very specific ones, which have very specific boundary graphs $\Gamma$. Furthermore, we do not consider the whole range of the sum (\ref{Eq:SF_StateSum}), but restrict ourselves to very specific states, which we will introduce in the following section \ref{Sec:Appr:QuantumCuboids}. This obviously changes the sum in and of itself, but we can expect the result to be still a somewhat  decent approximation in all those cases, in which the full sum is actually dominated by the states we consider. It seems reasonable to assume that, our choice of states -- the so-called \emph{quantum cuboids} -- dominate the path integral when describing flat space. It is still difficult, at this point, to estimate the error made by this approximation. For now, this question has to remain open, but we hope to return to it in a future article.

\item Secondly, we do not hope to be able to solve (\ref{Eq:RenormalizedAmplitude_abstract}) for generic $\mathcal{A}_\Gamma$, but rather search for solutions in the specific set of the EPRL-FK model, with Barbero-Immirzi parameter $\gamma<1$. In particular, we will restrict the flow to the space of EPRL-FK amplitudes with coupling constants $\gamma,\alpha$, by projecting the amplitude after each RG step. The details of this will be described in section \ref{Sec:Appr:FlowTruncation}.

\item Thirdly, we will work in the large-$j$-approximation of the model. This will have the advantage of having an explicit expression for the amplitude in terms of its boundary state spins. This formula was derived and investigated in \cite{Bahr:2015gxa}. We will give a recap of this analysis in appendix \ref{Sec:Appx:Amplitude}.
\end{itemize}

\subsection{Quantum cuboids}\label{Sec:Appr:QuantumCuboids}

\begin{figure}[hbt!]
\includegraphics[width=0.3\textwidth]{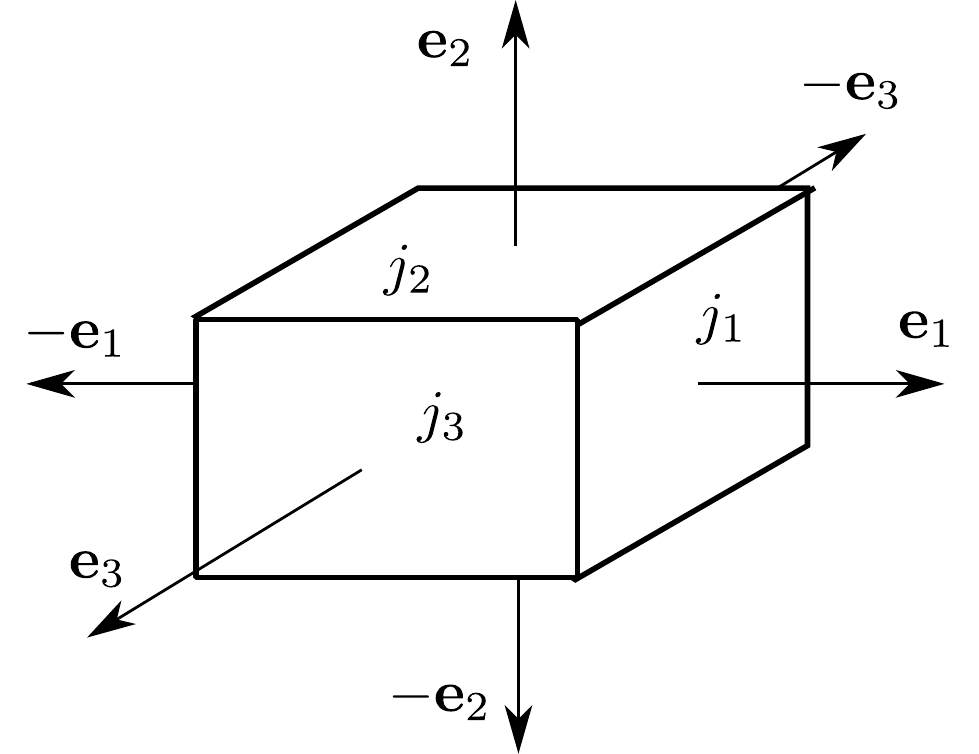}
\caption{A classical cuboid, the data of which is used to define the quantum cuboid intertwiners $\iota_e$. The $j_i$ are the three areas, by which the cuboid is uniquely defined up to rotation and translation. }\label{Fig:QuantumCuboid}
\end{figure}

The 2-complex we consider is determined by a $4D$ hypercubic lattice. Hence, the boundary graphs $\Gamma$ are regular, cubic graphs (apart from the corners), with six-valent nodes. In our symmetry-restricted model, we do not consider $\mathcal{H}_\Gamma$ to contain all spin network functions on $\Gamma$. Rather, we consider only those states which adhere to the cuboidal symmetry. In other words, we restrict ourselves to states for which the intertwiner $\iota_n$ at the nodes $n\in\Gamma$ are quantum cuboids. These are special cases of quantum polyhedra, which can be described by Livine-Speziale-intertwiners \cite{Livine:2007vk, Bianchi:2010gc}. A quboid with three face areas $j_1,j_2,j_3$, is described by the state

\begin{eqnarray}\label{Eq:CoherentCuboidIntertwiner}
\left | \iota_{n} \right \rangle = \int_{\text{SU}(2)} dg\;g\triangleright\bigotimes_{i = 1}^3 \left | j_i \, {\rm\bf e}_i \right \rangle \otimes \left\langle j_i\, -\!{\rm\bf e}_i \right |,
\end{eqnarray}

\noindent where ${\rm\bf e}_i$ are the three unit vectors along the three coordinate axes in $\mathbbm{R}^3$. This corresponds, in the large $j$-limit, to a polyhedron in the shape of a cuboid, see figure \ref{Fig:QuantumCuboid}.

The Hilbert spaces $\mathcal{H}_\Gamma$ we use consist of only the subspace of all spin network functions with intertwiners of the form (\ref{Eq:CoherentCuboidIntertwiner}). Such states are completely determined by the spin assignment $j_l$ to links $l$ of the boundary graph $\Gamma$. Note that, as a consequence, links which are on opposite ends of a node must have coinciding spins. Since this must hold for all nodes in the boundary graph, the allowed states are quite restricted in their form. In particular, fluctuations of a spin on a link $l$ are automatically translated in the direction parallel to that link. Fluctuations are therefore only local in directions orthogonal to $l$.

\begin{figure}[hbt!]
\includegraphics[width=0.3\textwidth]{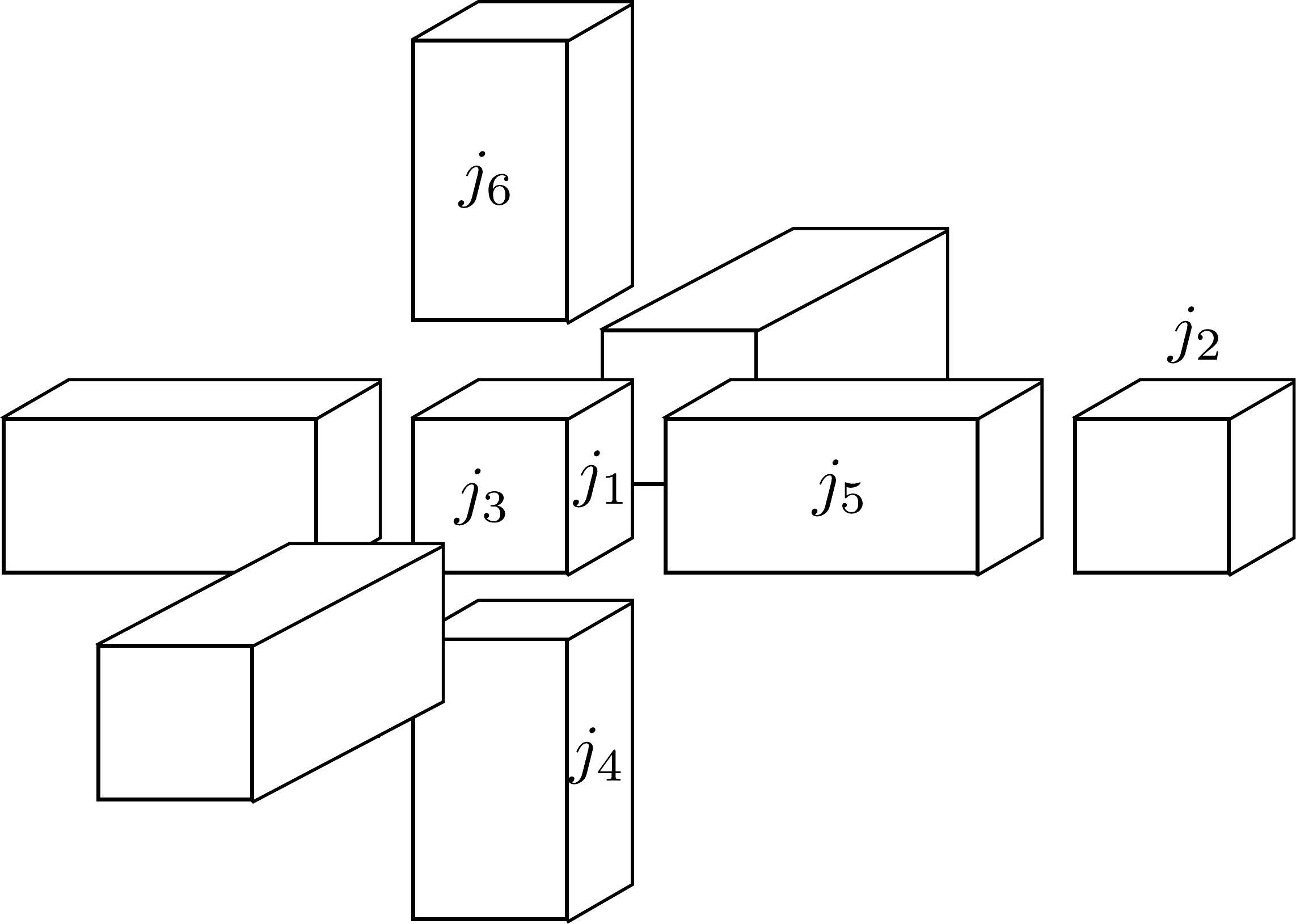}
\caption{The boundary of a hypercuboid in $4D$ consists of eight cuboids. The vertex amplitude associated to this hypercuboid depends on six spins $j_1,\ldots,j_6$. }\label{Fig:Hypercuboid_01}
\end{figure}

In a large $4D$ hypercubic lattice, a vertex $v$ does therefore correspond to a $4D$ hypercuboid, the boundary of which is given by eight quantum cuboids (see figure \ref{Fig:Hypercuboid_01}). Due to the symmetry of the cuboids, there are six independent spins in the boundary of one vertex.

\subsection{Cuboidal embedding maps}\label{Sec:Appr:Embmaps}

Now that we have defined $\mathcal{H}_\Gamma$ for all lattices $\Gamma$, it is important to specify how states on different boundary graphs $\Gamma$ interact. For this we need to choose the embedding maps $\iota_{\Gamma\Gamma'}$ (\ref{Eq:EmbeddingMaps}).

In principle, there are several possible choices for the $\iota_{\Gamma\Gamma'}$. It seems reasonable that one should choose the embedding maps such that the geometric interpretation of boundary states is respected.\footnote{Generally, it would be desirable to choose dynamical embedding maps \cite{Dittrich:2013xwa}, which are such that the additional degrees of freedom that one encounters by going over from $\Gamma'$ to $\Gamma$ are all in the physical vacuum state. However, this requires a great deal of knowledge about the dynamics of the system. Also, given that, depending on the phase space structure and the RG flow, there might be different fixed points corresponding to different (physical) vacua, we refrain from going this route. Instead, we choose embedding maps which respect the geometric interpretation of the boundary states, in the large $j$ limit. While this is not the only choice, it is one which presents itself by looking at the path integral (\ref{Eq:SF_StateSum}).}

\begin{figure}[hbt]
\includegraphics[width=0.35\textwidth]{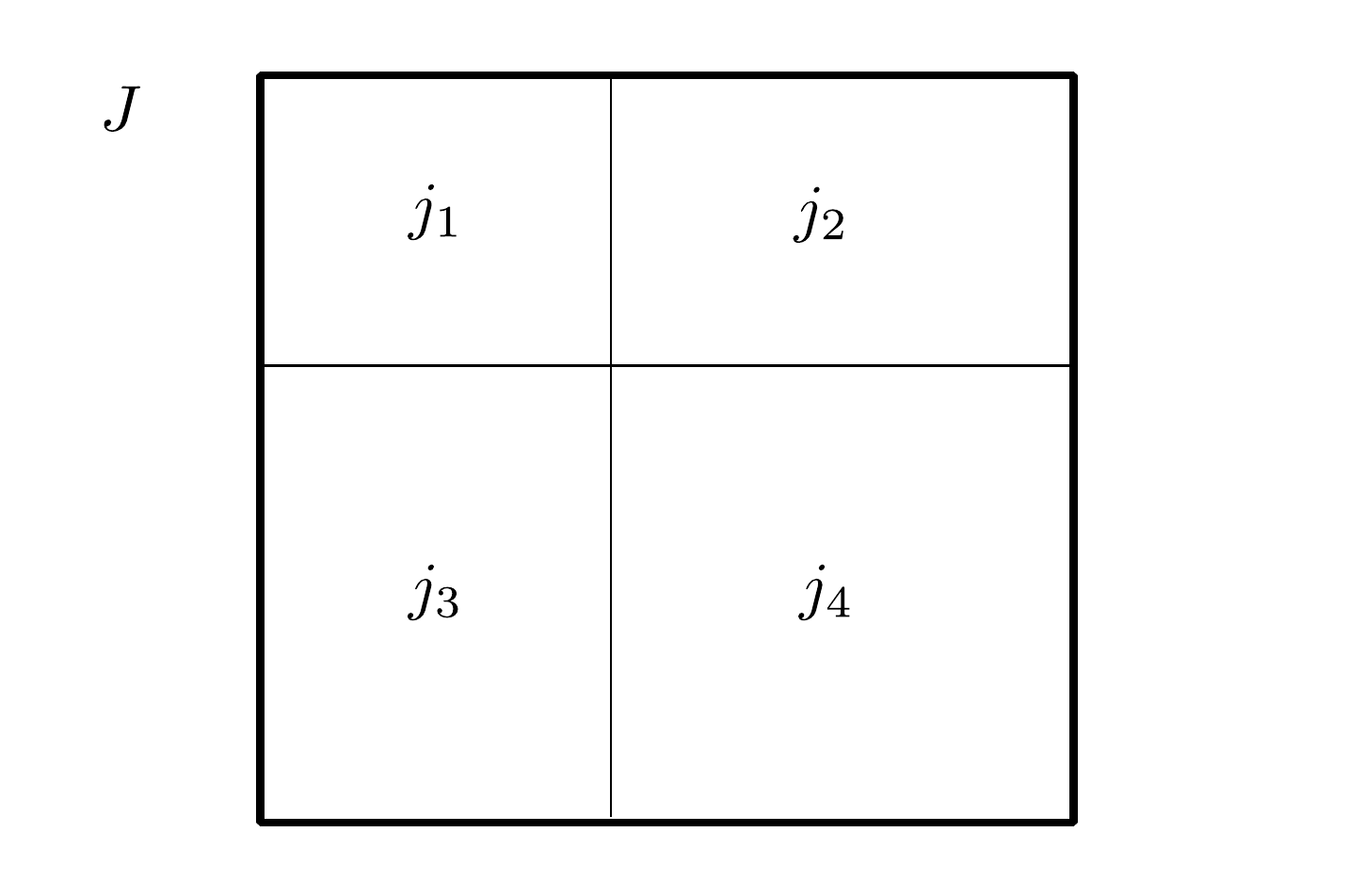}
\caption{In the coarse-graining step of the hypercubic lattice, each coarse face $F$ (large rectangle) consists of four fine faces $f$ (smaller rectangles). The embedding map rests on the fact that areas add up, i.e.~$J=j_1+\cdots j_4$. }\label{Fig:figure_Face_SubdividedHypercubic}
\end{figure}

Consider a (quantum cuboid) state $\psi^{(\Gamma')}_{\vec J}$ on the coarse lattice $\Gamma'$. From a geometric point of view, it seems reasonable that $\iota_{\Gamma\Gamma'}$ should be chosen such that $\iota_{\Gamma\Gamma'}\psi^{(\Gamma')}_{\vec J}$ is a linear superposition of  $\psi^{(\Gamma)}_{\vec j}$ which are as close as possible to $\psi^{(\Gamma')}_{\vec J}$, in terms of geometric observables measurable on the coarse graph $\Gamma'$. The most obvious observables are dihedral angles and areas. Since dihedral angles in all quantum cuboids states are $\frac{\pi}{2}$, only the areas remain. So, we choose
\begin{eqnarray}\nonumber
\iota_{\Gamma\Gamma'}\psi_{\vec J}^{(\Gamma')}\;=\;\frac{1}{N_{\vec J}}\sum_{j_e}\left(\prod_{\text{squares }E}\delta\big(J_E-\sum_{e\subset E}j_e\big)\right)\;\psi_{\vec j}^{(\Gamma)},\\[5pt]\label{Eq:EmbeddingMapQuboid}
\end{eqnarray}

\noindent where the normalization factor $N_{\vec J}$ is computed in section \ref{Sec:Appx:Normalization}, and the sum ranges over all fine quantum cuboid spin network functions which satisfy that, for each coarse square $E$ in the dual lattice, the four fine spins $j_e$ associated to squares $e$ that comprise it, add up to $J_E$ (see figure \ref{Fig:figure_Face_SubdividedHypercubic}).

In other words, the states $\psi_{\vec J}^{(\Gamma')}$ and $\iota_{\Gamma\Gamma'}\psi_{\vec J}^{(\Gamma')}$ coincide on the observables for the coarse areas $\hat{\text{Ar}}_E$.
It should be noted that, in principle, one could choose different $\iota_{\Gamma\Gamma'}$. Geometrically, however, this seems to be the most natural choice.

Using the hypercuboidal symmetry of the lattice, and the fact that the quantum cuboid intertwiners are completely determined by their spins, the restricted spin foam state sum (\ref{Eq:SF_StateSum}) can be written as
\begin{eqnarray}\label{Eq:DressedAmplitudeStateSum}
Z\;=\;\sum_{j_f}\prod_v\hat{\mathcal{A}}_v,
\end{eqnarray}

\noindent where face- and edge amplitudes have been absorbed into the dressed vertex amplitude $\hat{\mathcal{A}}_v$.

\subsection{Renormalization of the amplitude}\label{Sec:Appr:FlowTruncation}

In this article we operate with the large-$j$-asymptotics of the hypercuboidal amplitude, which all depend on the coupling constant $\alpha$.\footnote{Remember that, in the large $j$ asymptotics, the dependence of $\gamma$ vanishes for the quantum hypercuboids. See also the discussion in \cite{Bahr:2015gxa}.} In general, it will not be possible to find solutions to (\ref{Eq:RenormalizedAmplitude_abstract}) if one restricts to only those amplitudes, i.e.~for a given $\alpha$ there might \emph{not} be an $\alpha'$ such that
\begin{eqnarray}\label{Eq:RenznWishfulThinking}
\mathcal{A}_{\Gamma'}^{(\alpha')}\;=\;\mathcal{A}_\Gamma^{(\alpha)} \iota_{\Gamma\Gamma'},
\end{eqnarray}

\noindent holds. Therefore we resort to a strategy for the computation of the RG flow which is very popular also in other renormalization attempts, such as the asymptotic safety scenario (see \cite{Niedermaier:2006wt} and references therein): we truncate the flow to the space of hypercuboidal EPRL-FK amplitudes, by projecting the result of the computation down to it after every step.

This will be achieved using a procedure which, in this context, has been described in \cite{Bahr:2016hwc}: We first compute the renormalized amplitude $\mathcal{A}{(\text{ren})}_{\Gamma'}$ by
\begin{eqnarray}
\mathcal{A}^{(\text{ren})}_{\Gamma'}\;:=\;\mathcal{A}_\Gamma^{(\alpha)}\iota_{\Gamma\Gamma'}.
\end{eqnarray}

\noindent Of course, there is no guarantee that $\mathcal{A}^{(\text{ren})}_{\Gamma'}$ can be written exactly as $\mathcal{A}^{(\alpha')}_{\Gamma'}$ for some $\alpha'$, since additional couplings are usually generated during coarse graining. We therefore, as a second step in the RG procedure, \emph{project} this amplitude to the space of hypercuboidal amplitudes. Such a projection can in principle be achieved by many different means, which could lead to slightly different RG flows.\footnote{There is an example discussed in \cite{Bahr:2012qj}, in which a fixed point of the RG flow can vanish with a particularly bad choice of projection.} We choose the method described in \cite{Bahr:2016hwc}, by choosing a reference observable $O$, and define the renormalized value of $\alpha'$ to be the number for which the corresponding hypercuboidal amplitude and the renormalized amplitude agree the most. In other words, for some boundary state $\psi_{\Gamma'}$, we define the renormalized coupling constant $\alpha'$ to be such that
\begin{eqnarray*}
\Delta_{O,\alpha,\alpha',\psi_{\Gamma'}}\;:=\;\left\|\Big(\mathcal{A}^{(\text{ren})}_{\Gamma'}-\mathcal{A}^{(\alpha')}_{\Gamma'}\Big)O\psi_{\Gamma'}\right\|^2\;\stackrel{!}{=}\;\text{min}
\end{eqnarray*}

\noindent is minimal. \footnote{In principle, one could choose an orthonormal basis (ONB) of $\mathcal{H}_{\Gamma'}$ and a complete set of observables $O$, and demand that the sum -- assuming it exists -- of the errors is minimal. This would be equivalent to demanding that the Hilbert space norm of $\mathcal{A}^{(\text{ren})}_{\Gamma'}$ and $\mathcal{A}^{(\alpha')}_{\Gamma'}$, seen as vectors in $\mathcal{H}_{\Gamma'}$ (via the Riesz representation theorem), is minimal. In other words, this would actually be a projection with respect to the boundary Hilbert space inner product.}

The observable should, ideally, be chosen such that it can resolve the coupling constant $\alpha$. In other words, one should choose a coupling constant from which one can infer $\alpha$, i.e.~such that the map $\alpha'\mapsto \langle O\rangle=\mathcal{A}^{(\alpha')}(O\psi_{\Gamma'})$ is injective, at least in the region in which one suspects the renormalized coupling constant to lie.

\begin{figure}[hbt]
\includegraphics[width=0.45\textwidth]{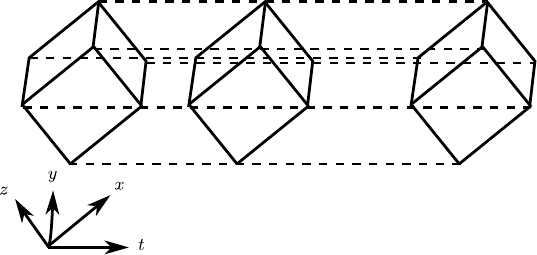}
\caption{Two hypercuboids, glued together at a space-like cuboid, in the $xyz$-hyperplane.}\label{Fig:figure_VertexDisplacementHypercuboids}
\end{figure}

\begin{figure}[hbt]
\includegraphics[width=0.45\textwidth]{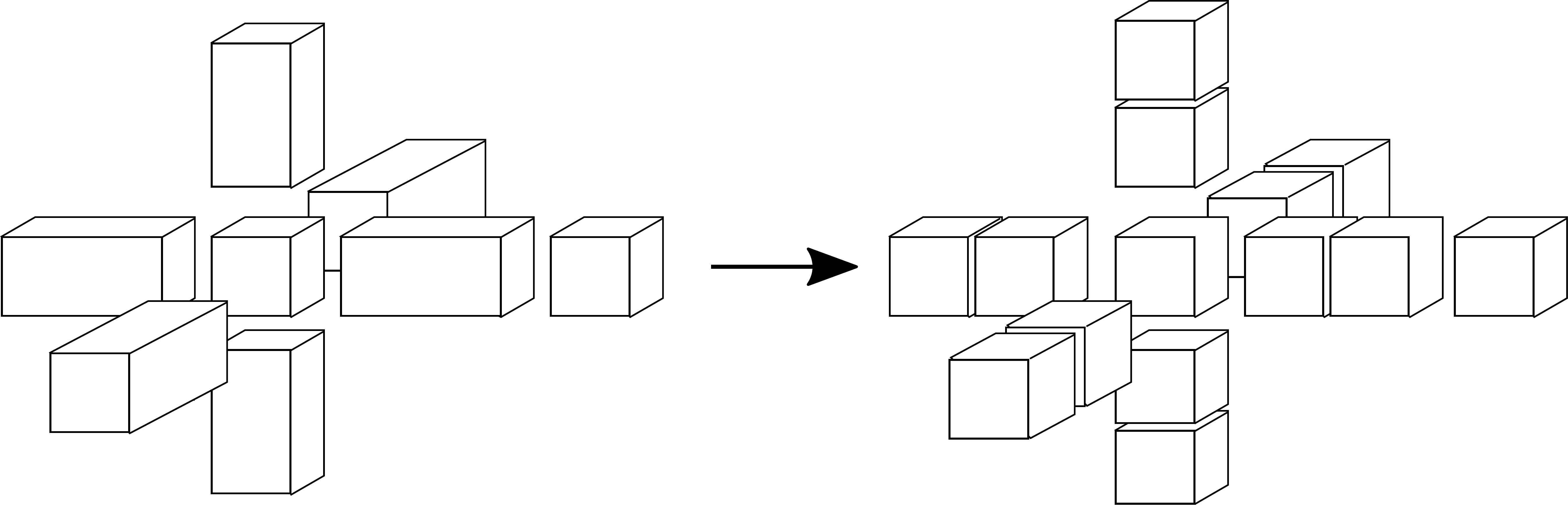}
\caption{Subdivision of one hypercuboid into two hypercuboids, glued together at a space-like cuboid, in the $xyz$-hyperplane. The right figure is the dual to the boundary graph of the lattice depicted in figure \ref{Fig:figure_VertexDisplacementHypercuboids}}\label{Fig:figure_Hypercube_SplitInTwo}
\end{figure}

A good choice for this observable can be inferred from the analysis in \cite{Bahr:2015gxa}: it is the fluctuation of the $4$-dimensional volume. Consider a lattice consisting of only \emph{two} hypercuboids $v_1$, $v_2$, which are glued together along a common (spatial) cuboid (see figure \ref{Fig:figure_VertexDisplacementHypercuboids}). As boundary state we choose $\psi_{\Gamma'}:=\iota_{\Gamma'\Gamma''}\psi_{\Gamma''}$, where $\Gamma''$ is the boundary of the single hypercuboid $V$ which consists of $v_1$ and $v_2$ (see figure \ref{Fig:figure_Hypercube_SplitInTwo}), and $\psi_{\Gamma''}$ being some boundary state. The choice for $O$ is then
\begin{eqnarray}\label{Eq:Observable}
O\;=\;\Delta\mathcal{V}_1\;=\;\Big(\mathcal{V}_1-\langle \mathcal{V}_1\rangle\Big)^2.
\end{eqnarray}

\noindent It is not difficult to see that the expectation value of $O$ corresponds to the fluctuation of the volume $\mathcal{V}_1$ of one of the hypercuboids. Of course, given the embedding map, the sum ranges over states $j_f$ for which the total volume $\mathcal{V}_1+\mathcal{V}_2\equiv\mathcal{V}$ (which is determined by the boundary state $\psi_{\Gamma''}$) is constant. For symmetry reasons, obviously $\langle \mathcal{V}_1\rangle=\frac{1}{2}\mathcal{V}$ is just half of the total volume, no matter what the value of $\alpha$. However, the fluctuation of the volume increases strictly monotonically as $\alpha$ decreases \footnote{See e.g.~figure \ref{Fig:figure_vts01} in the following section. This is, of course, directly connected to the interplay between the value of $\alpha$ and the vertex translation symmetry of the amplitude.}. With this bit of knowledge, we can perform the RG step, by computing the fluctuation of the volume of the renormalized amplitude. Since $\alpha'\mapsto \langle\Delta\mathcal{V}_1\rangle$ is invertible (in the region we are interested in), we can infer the renormalized value of $\alpha'$.\footnote{Note that, since we have as many reference observables as coupling constants, we can find a renormalized value such that $\Delta_{O,\alpha,\alpha',\psi_{\Gamma'}}=0$ in this case.}

\section{Results}\label{Sec:Results}

In the following, we consider different results of our renormalization computations.

\subsection{Vertex translations}\label{Sec:Results:VTS}

By vertex translations we mean a transformation of the variables $j_f$, which corresponds to a diffeomorphism on the lattice. They have been described in detail in \cite{Bahr:2015gxa}. On a geometric configuration a vertex translation can be most easily visualized by a translation of a 3d hyperplane in its orthogonal direction. For two hypercuboids $v_1$, $v_2$ meeting at two of their cuboids, their respective spins are $j_i$ and $k_i$, with $i=1,\ldots, 6$, satisfy $j_i=k_i$ for $i=1,2,3$. The action of the vertex translations are then given by
\begin{eqnarray}
\begin{array}{rcl}j_i\;&\to&\;(1+x)j_i,\\[5pt]
k_i\;&\to&\;(1-x)k_i
\end{array}
\qquad i=4, 5, 6.
\end{eqnarray}

\noindent It is clear that, for a setup in which $j_i=k_i$ also for $i=4, 5, 6$, the function
\begin{eqnarray}\label{Eq:AmplitudeVTS01}
I^\alpha(x)\;:=\;\frac{\hat{\mathcal{A}}^{(\alpha)}_v(j_{r},(1+x)j_{s})\hat{\mathcal{A}}^{(\alpha)}_v(j_{r},(1-x)j_{s})}{\hat{\mathcal{A}}^{(\alpha)}_v(\vec{j})^2}
\end{eqnarray}

\noindent (with $r=\{1,2,3\}$ and $s=\{4,5,6\}$) satisfies $I^\alpha(0)=1$ and $(I^\alpha)'(0)=0$, for any value of $\alpha$. This is simply a consequence of the symmetry of the situation. However, in general $I^{\alpha}(x)$ will not be constant for $x\in[-1,1]$, meaning that not all contributions which arise as vertex translations of each other are weighted equally by the path integral.

However, one finds that there is a specific value of the coupling constant $\alpha$ for which $I^\alpha(x)$ indeed is nearly constant among the whole range of $x$, apart from the very boundary of the interval $[-1, 1]$. At these points, however, some of the spins become rather small, so the large-$j$-asymptotic formula can not be trusted completely anyway at these points.

Choosing boundary data which is \emph{geometric}, i.e.~which satisfies the geometricity constraints
\begin{eqnarray}\label{Eq:GeometricityConstraint}
j_1j_6=j_2j_5=j_3j_4,
\end{eqnarray}
\noindent we can go over from spins (areas) to edge lengths, i.e.~four numbers $X, Y, Z, T$ describing the edge lengths of the two hypercuboids. In other words
\begin{eqnarray}\nonumber
j_1&=&YZ,\quad j_2=XY,\quad j_3=XZ\\[5pt]\nonumber
j_4&=&YT,\quad j_5=ZT,\quad j_6=XT,
\end{eqnarray}

\noindent as can be inferred from figures \ref{Fig:Hypercuboid_01} and \ref{Fig:figure_VertexDisplacementHypercuboids}.

\begin{figure}[hbt]
\includegraphics[width=0.45\textwidth]{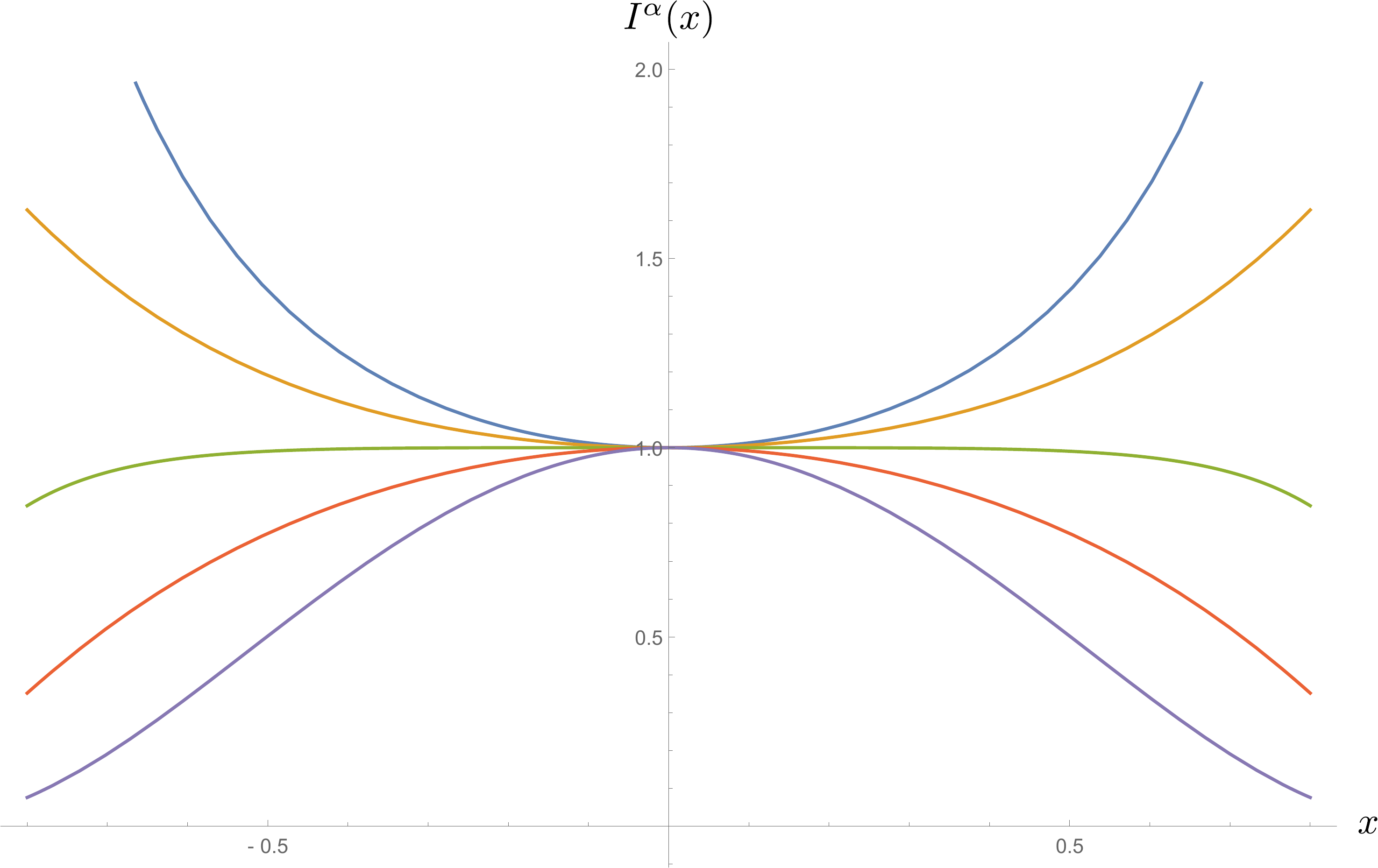}
\caption{Action of the vertex translations: depicted are $I^\alpha(x)$, depending on $x$, for various values of $\alpha$, for the boundary data $X=Y=Z=1$, $T=1$.}\label{Fig:figure_vts01}
\end{figure}

\begin{figure}[hbt]
\begin{center}
\includegraphics[width=0.45\textwidth]{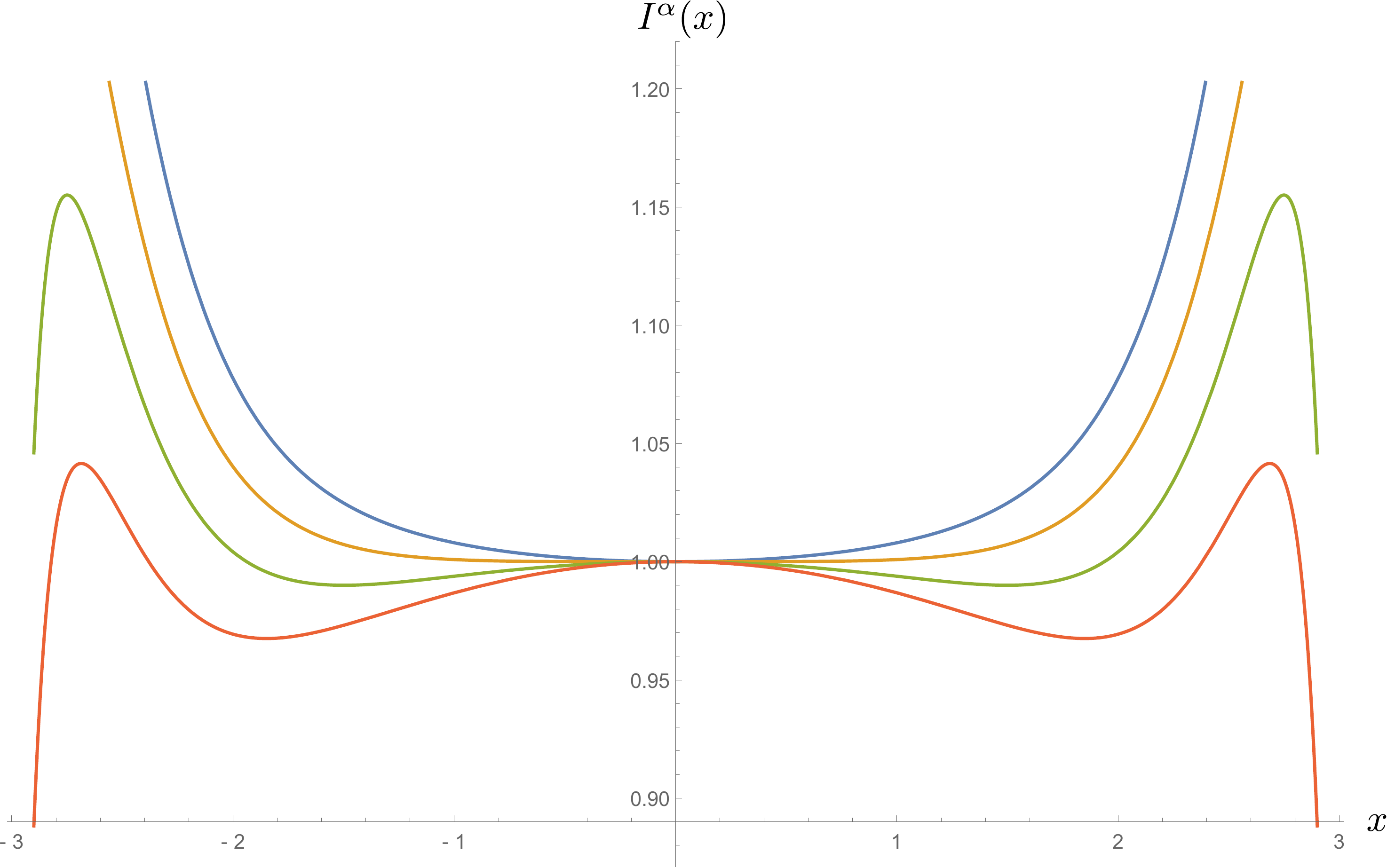}
\caption{Action of the vertex translations: depicted are $I^\alpha(x)$, depending on $x$, for various values of $\alpha$, for the boundary data $X=Y=Z=1$, $T=3$.}\label{Fig:figure_vts02}
\end{center}
\end{figure}

The behaviour of $I^\alpha(x)$ is shown for different boundary data $j_i$ and different $\alpha$ in figures \ref{Fig:figure_vts01}, \ref{Fig:figure_vts02}. The critical value $\alpha_c$, for which the amplitude becomes (nearly) invariant under vertex translations can be determined by solving $(I^{\alpha_c})''(x)_{|x=0}=0$ for $\alpha$. We have listed some values for the critical $\alpha$ for different boundary data.

\begin{center}
\begin{tabular}{l|c}\label{Tab:Tabular01}
boundary data $\qquad$ & $\qquad \qquad \alpha_c\qquad \qquad$\\[5pt]\hline
{}&{} \\
$X=Y=Z=T=1$ & $0.606667$\\[5pt]
$X=Y=Z=1$, $T=3$ & $0.549942$\\[5pt]
$X=3$, $Y=Z=T=1$ & 0.557808\\[5pt]
$X=3$, $Y=5$, $Z=T=1\quad$& 0.539846
\end{tabular}
\end{center}
\noindent As one can see, the critical value for $\alpha_c$ lies in a range of $0.54 - 0.61$, depending on the boundary data, i.e.~on the observable. It is noteworthy that the maximum value appears to lie at the point where the boundary is given by two completely regular coarse hypercuboids.

\subsection{Renormalization group step}

We come to the main part of the article: the computation of an RG step, using the methods outlined in section \ref{Sec:Renzn}. To reiterate: We start out with the hypercuboidal amplitude in the large $j$-approximation $\hat{\mathcal{A}}_v^{(\alpha)}$, depending on six spins $j_i$. We then define the renormalized amplitude $\hat{\mathcal{A}}^{(\text{ren})}$, which depends on six spins $J_i$, to be
\begin{widetext}
\begin{eqnarray}\label{Eq:RenormalizedAmplitude}
\hat{\mathcal{A}}^{(\text{ren})}(\vec J)\;=\;\frac{1}{N_{\vec J}}\int d^{24}j\;\prod_F\left(J_F-\sum_{f\subset F}j_f\right)\prod_{v=1}^{16}\hat{\mathcal{A}}_v^{(\alpha)}(\vec j)
\end{eqnarray}
\end{widetext}

\noindent Note that we make the coarse graining step from $2\times 2\times 2\times 2$ fine to one coarse hypercuboid here, which means that each of the six coarse faces $F$ is subdivided into four fine faces $f$. Since we are working in the large $j$-approximation, we replaced the sums over the $j$ by integrals. The delta functions and the normalization factor $N(\vec J)$ come from the embedding maps, which have been discussed in section \ref{Sec:Renzn:EmbMaps} and appendix \ref{Sec:Appx:Normalization}.

Solving for the delta functions in the integral, an integration over 18 spins remains in order to compute the renormalized amplitude (\ref{Eq:RenormalizedAmplitude}). This goes up to 33 spins when attempting to compute the RG step, which relies on comparing the expectation value of the observable (\ref{Eq:Observable}) on two coarse hypercuboids and $2\times 2\times 2\times 4=32$ fine hypercuboids, for fixed coarse boundary spins.

Since this is numerically quite involved, we make our first computation by restricting ourselves to the \emph{geometric sector}, i.e.~the subset of spins such that, in each hypercuboid, the geometricity constraints (\ref{Eq:GeometricityConstraint}) are satisfied (using the notation of figure \ref{Fig:Hypercuboid_01}). For values of $\alpha$ which are not too small, this is indeed a good approximation, as non-geometric configurations are suppressed for high values of $\alpha$, as has been shown in \cite{Bahr:2015gxa}. \footnote{Indeed, one could argue that these non-geometric configurations, which are similar to twisted geometries \cite{Freidel:2010aq}, arise due to a wrong implementation of the volume simplicity constraint on the hypercuboidal amplitude \cite{Perez:2012wv}. A correct implementation of this constraint should remove these from the path integral.}

On geometric configurations, the spins for each vertex can be reverted to four edge lengths, which define a $4D$ hypercuboid. This leaves us with the integration over one variable on the coarse lattice (the length of one of the hypercuboids in the $T$-direction), and $6$ variables on the fine lattice (one each for the $X$-, $Y$- and $Z$-direction, three for the $T$-direction.

One should be careful to note that, when going over from spins $j=l_1l_2$ to lengths $l_i$, there is a nontrivial Fadeev-Popov-determinant, which has to be included in the path integral. We derive this in appendix \ref{Sec:Appx:FadeevPopov}.

We have carried out the integrals numerically, for several different boundary values $X, Y, Z, T$. See appendix \ref{Sec:Appx:Numerical} for details on the employed methods.

\begin{figure}[hbt]
\includegraphics[width=0.45\textwidth]{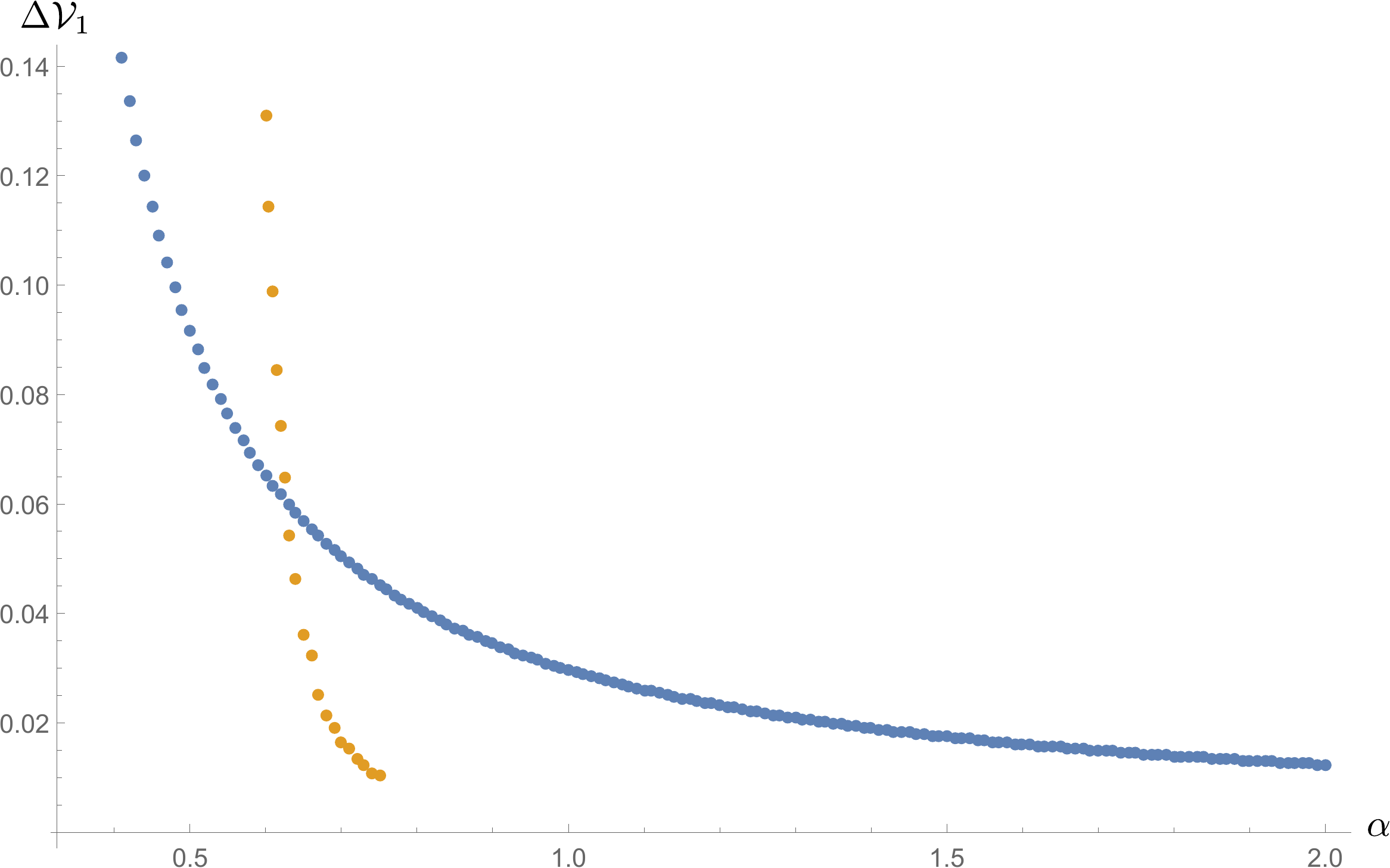}
\caption{Expectation value of the observable (\ref{Eq:Observable}, depending on $\alpha$. The boundary data is $X=Y=Z=T=1$, and we use the hypercuboidal amplitude (blue), and the renormalized amplitude (orange). The two intersect at $\alpha^*\approx 0.628$.)}\label{Fig:image_ObservablesBdy01}
\end{figure}

Figure \ref{Fig:image_ObservablesBdy01} shows the expectation value of our chosen observable (\ref{Eq:Observable}) depending on $\alpha$, for a given (coarse) boundary state, $X=Y=Z=T=1$, once for the original (blue), and once for the renormalized (orange) amplitude. As one can see, there is an intersection between the two curves which indicates that for this particular value of $\alpha=\alpha^*$, the two amplitudes agree on the observable (\ref{Eq:Observable}). Since we use this observable to define our flow, this is also the RG fixed point.

\begin{figure}[hbt]
\includegraphics[width=0.45\textwidth]{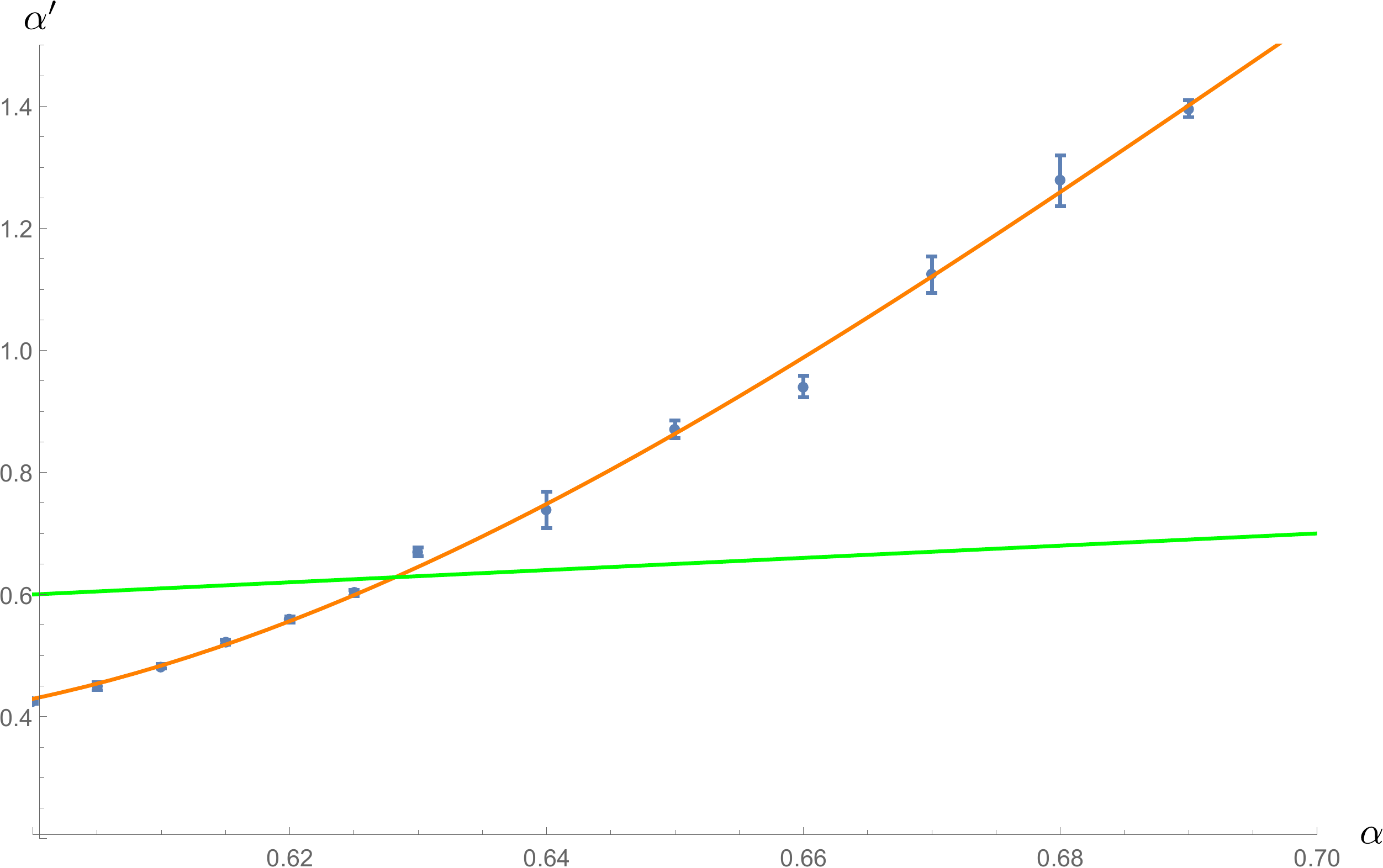}
\caption{Renormalized value of $\alpha'$ depending on $\alpha$ (orange), together with numerical fit, for boundary data is $X=Y=Z=T=1$. Iterating this function generates the RG flow $\alpha\to\alpha'\to\alpha''\to\ldots$. The intersection with the line $\alpha'=\alpha$ (green) indicated the RG fixed point.}\label{Fig:image_RGFlow_Bdy01}
\end{figure}

The explicit flow, given by the map $\alpha\mapsto\alpha'$, is depicted in figure \ref{Fig:image_RGFlow_Bdy01}. The fixed point there is given by the intersection fo the graph of the curve with the line $\alpha=\alpha'$. We get
\begin{eqnarray}\label{Eq:FixedPoint01}
\alpha^*\;\approx\;0.628.
\end{eqnarray}

\noindent The fact that the slope of $\alpha\mapsto\alpha'$ is larger than $1$ at the intersection, indicates that the fixed point $\alpha^*$ of the RG flow is unstable. Since this flow goes from the fine to the coarse lattice, the point can be said to be UV attractive. However, we stress that ``UV'' in this setting is not defined in terms of some length scale, but with respect to the fine-ness of the used lattice.

One can repeat this analysis for several values of boundary data, which, in essence, amounts to changing the observable with respect to which the RG flow of $\alpha$ is defined. Remember that this is because the boundary state can be regarded as part of the observable on a lattice with boundary.

\begin{figure}[hbt]
\includegraphics[width=0.45\textwidth]{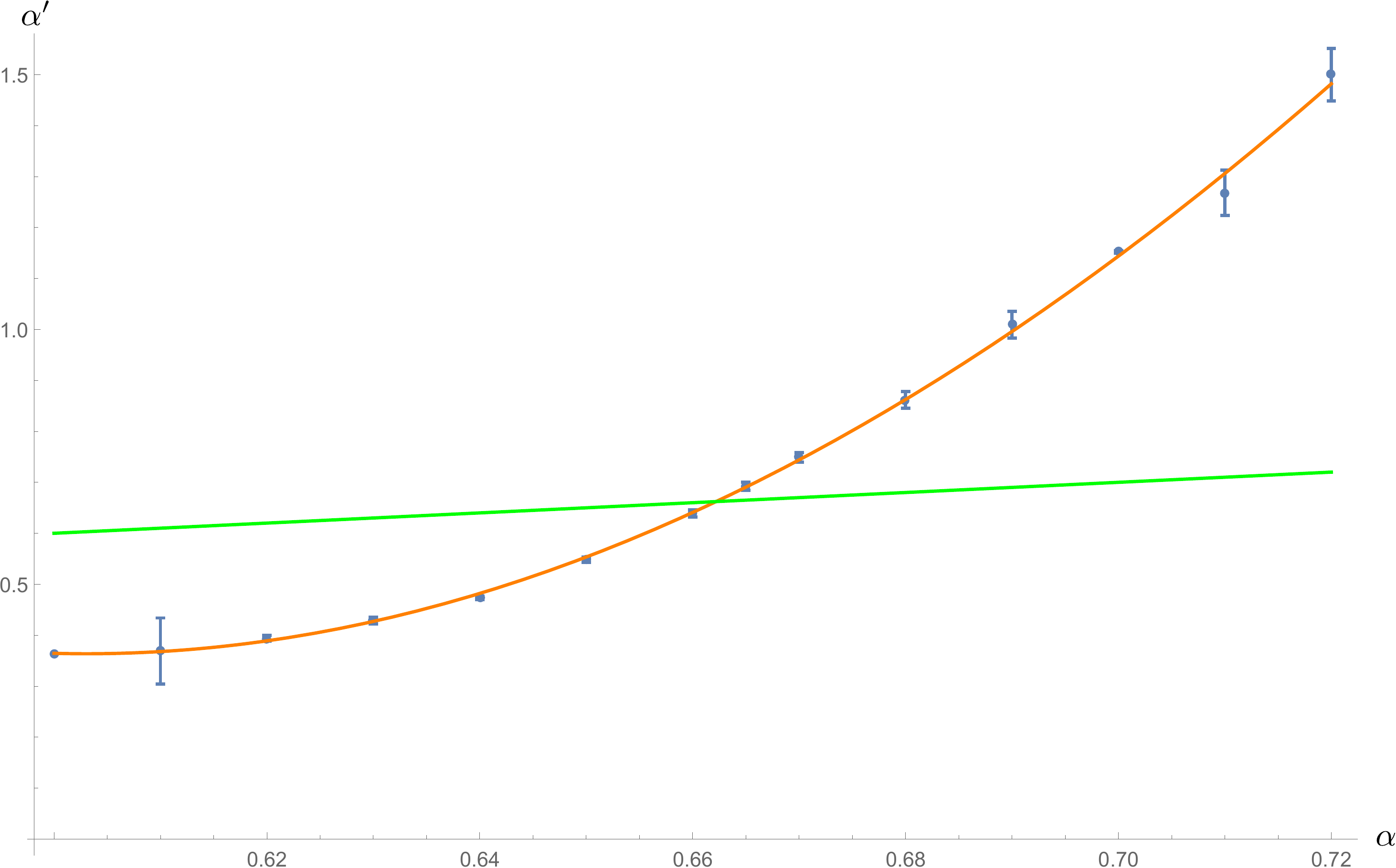}
\caption{RG step $\alpha\mapsto\alpha'$ for boundary data $X=Y=Z=1$, $T=3$ (orange). The intersection with $\alpha=\alpha'$ (green) lies at $\alpha^*=0.662$.}\label{Fig:image_RG-Step-results_bdry-1}
\end{figure}

\begin{figure}[hbt]
\includegraphics[width=0.45\textwidth]{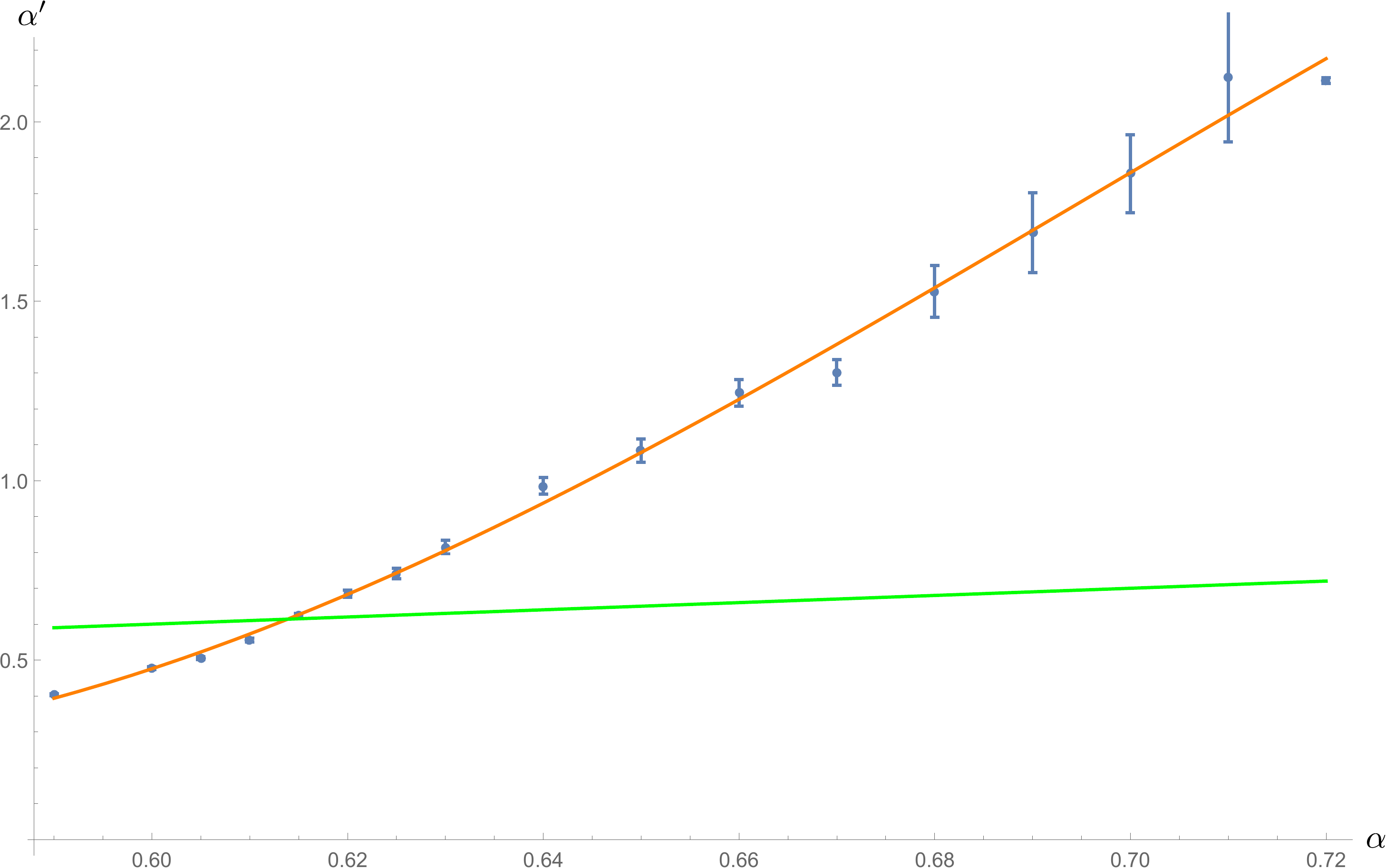}
\caption{RG step $\alpha\mapsto\alpha'$ for boundary data $X=3$, $Y=Z=T=1$ (orange). The intersection with $\alpha=\alpha'$ (green) lies at $\alpha^*=0.614$.}\label{Fig:image_RG-Step-results_bdry-2}
\end{figure}

\begin{figure}[hbt]
\includegraphics[width=0.45\textwidth]{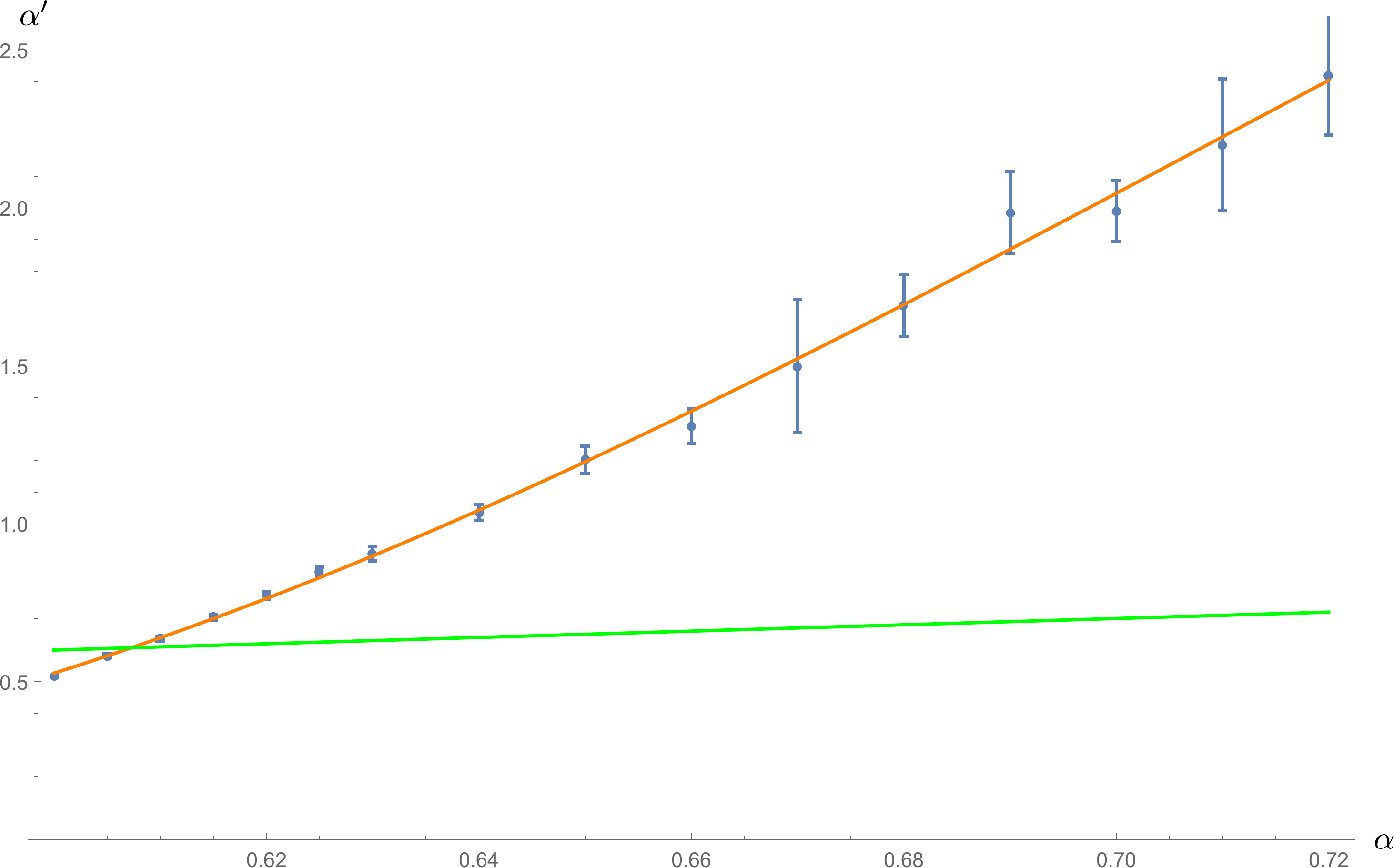}
\caption{RG step $\alpha\mapsto\alpha'$ for boundary data $X=3$, $Y=5$, $Z=T=1$ (orange). The intersection with $\alpha=\alpha'$ (green) lies at $\alpha^*=0.607$.}\label{Fig:image_RG-Step-results_bdry-3}
\end{figure}

In figures \ref{Fig:image_RG-Step-results_bdry-1}, \ref{Fig:image_RG-Step-results_bdry-2} and \ref{Fig:image_RG-Step-results_bdry-3}, we show numerical results for varying boundary states. As one can see, the situation is qualitatively similar in all cases. This supports the hope that the fixed point is, indeed, a characteristic feature of the model. The numerical value of the fixed point changes slightly, however, if the boundary data changes. This is not too surprising since, strictly speaking, changing the boundary state amount to changing the truncation of the RG flow, i.e.~the approximation in the RG calculation.

It should be noted that the relative numerical errors are in the order of magnitude below one percent. This is considerably lower than in the original article \cite{Bahr:2016hwc}, and mostly due to the much improved numerical methods employed in our analysis. See appendix \ref{Sec:Appx:Numerical} for details.

\subsection{Unprojected amplitude}

One major step in computing the RG flow was the projection of the renormalized amplitude (\ref{Eq:RenormalizedAmplitude}) to the subset of hypercuboidal amplitudes, which are parameterized by $\alpha$. As an investigation to how large the error is upon approximation, it is interesting to compare the renormalized amplitude $\hat{\mathcal{A}}^\alpha_\Gamma\iota_{\Gamma\Gamma'}$, and the projected renormalized amplitude $\hat{\mathcal{A}}^{\alpha'}_{\Gamma'}v$. Since both are functions of six (coarse) spins, there are several ways of comparison one could think about.

\begin{figure}[hbt]
\includegraphics[width=0.45\textwidth]{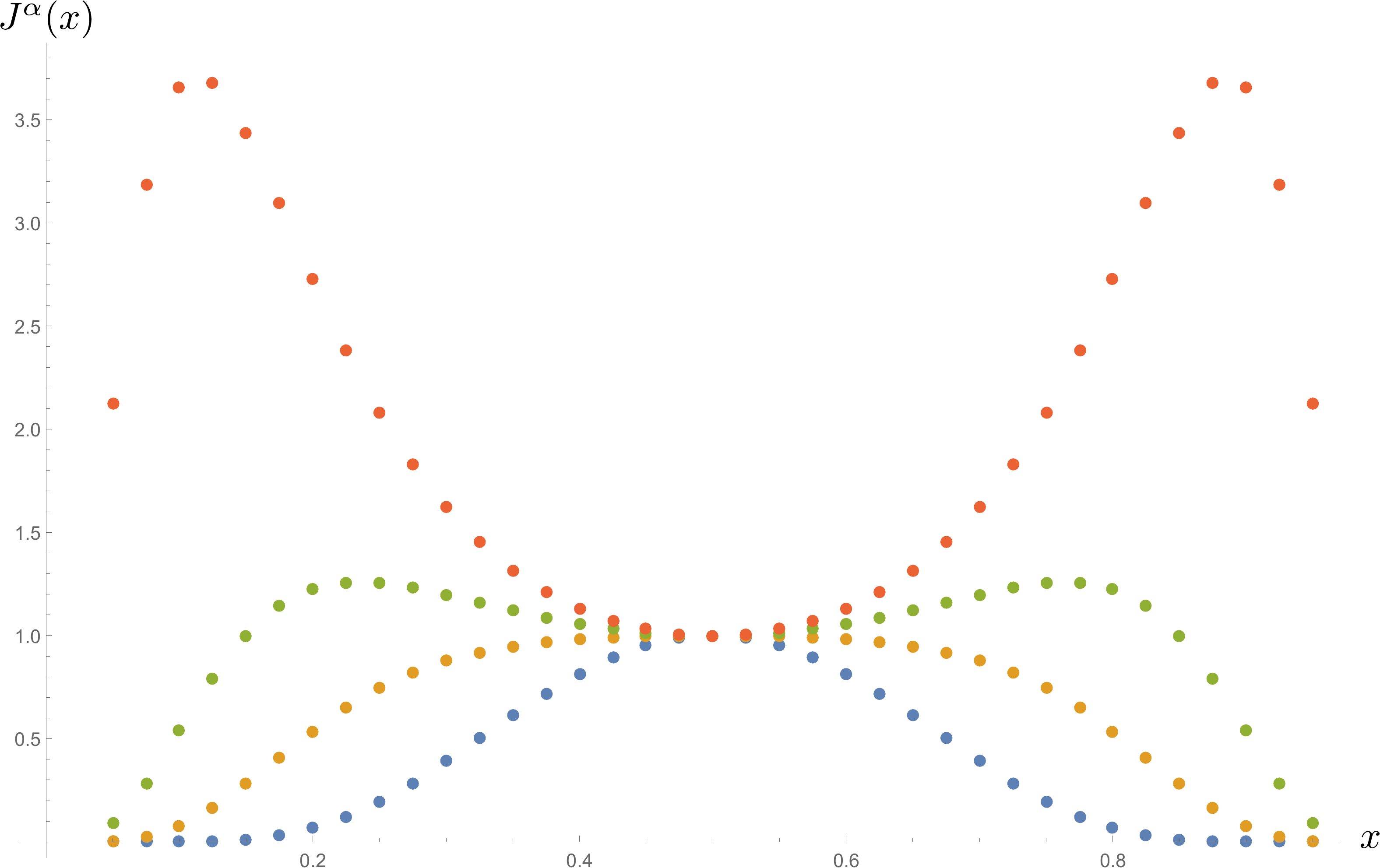}
\caption{Vertex translation symmetry of the renormalized, unprojected amplitude. Plotted is the value $J^\alpha(x)$, depending on $x$, for different values of $\alpha=0.61,0.63,0.65,0.70$. }\label{Fig:image_Ren-ampl-vertex-translation}
\end{figure}

In our previous analyses, one important property of the amplitudes was the behaviour under vertex translation symmetry. In particular, the value parameter $\alpha$ governed whether, in the path integral, regular or irregular subdivisions dominate. To this end, we investigate the situation of two hypercuboids glued together at a common spatial cuboid. This setup is identical to the one described in section \ref{Sec:Results:VTS}. See that section for more details. We plot the value
\begin{eqnarray}\label{Eq:VTS_RenAmpl}
J^\alpha(x)\;:=\;\frac{\hat{\mathcal{A}}^{\text{ren}}(J_i,2xK_i)\hat{\mathcal{A}}^{\text{ren}}(J_i,2(1-x)K_i)}{\hat{\mathcal{A}}^{\text{ren}}(J_i,K_i)^2}
\end{eqnarray}

\noindent for boundary data $J_i, K_i$ in terms of $X=Y=Z=T=1$. The result can be seen in figure \ref{Fig:image_Ren-ampl-vertex-translation}.

One can see nicely that the behaviour of the renormalized ammplitude is very similar to the projected renormalized amplitude: For small values of $\alpha$, the amplitude with very irregular subdivisions dominate, while for large $\alpha$ regular subdivisions, where the $4D$-volume is distributed evenly on the hypercuboids, dominate the path integral. There is a critical value of $\alpha$ where one case goes over to the other. This is precisely where the second derivative of $J^\alpha(x)$ vanishes at $x=1$, the place of regular subdivision. The critical value lies at
\begin{eqnarray}\label{Eq:FixedPoint02}
\alpha_c\;\approx\;0.63.
\end{eqnarray}

\noindent It is noteworthy that this value lies pretty close to the fixed point for the projected amplitude (\ref{Eq:FixedPoint01}), but slightly higher than the critical point of the hypercuboidal amplitude ($0.60667$ in table \ref{Tab:Tabular01}). Still, all of them lie well within the numerical range of different values obtained by different methods throughout this article. It can be concluded that, at least as the vertex translation symmetry is concerned, the renormalized and the projected renormalized amplitude behave quite similar, both qualitatively and quantitatively. Further checks are, of course, needed in order to establish a precise measure of how severe the projection changes the amplitude. We hope to come back to this in the future.

\section{Summary and discussion}\label{Sec:SummAndDiss}

In this article we have investigated the hypercuboidal renormalization of the $4D$ EPRL-FK spin foam amplitude for Riemannian signature. The main goal was to provide details to the computation reported on in \cite{Bahr:2016hwc}, and confirm the results, using different methods. We found an agreement, in particular we could confirm the existence of a fixed point in the renormalization group flow of the hypercuboidal truncation of the model. The fixed point is persistent under changing of boundary data, although its precise numerical value fluctuates slightly.

In all of the article, we worked with several approximations, which have been discussed in section \ref{Sec:Appr}. The most crucial one is certainly the truncation of the model to hypercuboids. In particular, the intertwiners $\iota_e$ were restricted to quantum cuboids, i.e.~to be of the form (\ref{Eq:CoherentCuboidIntertwiner}). This restriction results in a simplification, which made it possible to access the amplitude numerically. However, no curvature degrees of freedom are being summed over in the truncated path integral.

Still, it is important to point out that the truncation is not a toy model, but a subset of the full theory. In particular, the truncated path integral includes many important terms which also appear in the full path integral. All statements about which of these are dominant in the truncated sum will still be valid in the full model, so our analysis teaches us something about the full model as well. Still, at this point it is unclear whether the fixed point, and in particular its properties with regards to the vertex translation symmetry, are a genuine feature of the model, or an artefact of the truncation. This is a point which we will address in future articles.

\subsection{Summary of the results}\label{Sec:SummAndDiss:Summ}

First we investigated the properties of the amplitude under vertex translation symmetry. In particular, we were concerned with the path integral of two hypercuboids, which were glued together at a common, spatial cuboid. We then considered the path integral for a boundary state for which the spatial spins are being kept fixed, and the time-like spins were always such that the total $4D$ volume of space-time was kept constant. In this case all the occurring states can be related by vertex translations, as described in detail in \cite{Bahr:2015gxa}. From a semiclassical perspective, where vertex translations are a manifestations of the action of the diffeomorphism group \cite{FreidelLouapreDiffeo2002, Dittrich:2008pw, Bahr:2009ku}, all these states should contribute equally in a diffeomorphism-invariant theory. All states in the path integral arise as different ways to subdivide one hypercuboid into two along the $xyz$-hyperplane.

The model, however, fails to be invariant under this symmetry, for almost all values of $\alpha$. Irrespective of the boundary values, one can say that in general, for large value of $\alpha$, very regular subdivisions dominate the path integral. Conversely, for small values of $\alpha$, irregular subdivisions are dominant. There is a critical value of $\alpha=\alpha_c$ which separates these two regions, which we have identified as the point where the second derivative of the path integrand (\ref{Eq:AmplitudeVTS01}) with respect to the vertex translation symmetry vanishes (see figure \ref{Fig:figure_vts01}).

The exact value of $\alpha_c$ slightly varies with the boundary state. All values found lie in the range of $0.54 - 0.61$.

Next we have performed an actual renormalization group step. This amounted to coarse graining $2^4=16$ hypercuboids into one, and reabsorb the corrections into a changed (i.e.~renormalized) coupling constant $\alpha'$. The flow $\alpha\mapsto\alpha'$ was computed for various $\alpha$, and various boundary data. Similarly to the case of the vertex translation symmetry, we found that small values of $\alpha$ flow to even smaller ones, while high values of $\alpha$ flow to even higher ones. As a consequence, there is an unstable fixed point in the intermediate regime. This fixed point $\alpha^*$ lies in the range of $0.60 - 0.66$, depending on the boundary data.

Finally, we have also compared the renormalized amplitude without projection, i.e.~the one which is not projected down to the hypercuboidal subset of theories. In comparing the behaviour of the two with respect to the vertex translation symmetry, we found close agreement between the unprojected and the projected amplitudes. This is very reassuring, in that it suggests that by projection one does not obtain a large error. It is interesting to note that for the renormalized unprojected amplitude, the fixed point $\alpha^*$ and the critical point $\alpha_c$ are much closer to each other (both around 0.628), while for the projected amplitude there is about a $5\%$ difference between the two. Still, given the large amount of approximations we have employed throughout the analysis, it is remarkable that all of these values are so close to one another.

We can summarize the results of this article the following way:
\begin{itemize}
\item We have confirmed the existence of the critical point $\alpha_c$ (i.e.~the point where the vertex amplitude becomes almost invariant under vertex translations) for various boundary data. The value for $\alpha_c$ changes only slightly with varying the boundary state, and lies between $0.54 - 0.61$.
\item We have confirmed the existence of the fixed point $\alpha^*$ of the truncated RG flow, for various boundary data. Using much more involved numerical methods than in the original article \cite{Bahr:2016hwc}, we were able to bring the numerical error bars down considerably. The value for $\alpha^*$ changes only slightly with varying the boundary state, and lies between $0.60 - 0.66$.
\item During the projection to the hypercuboidal amplitudes, the properties of the amplitudes do not change much. This suggests that the RG flow does not lead the theories away from the hypercuboidal subset of theories in a significant way.
\end{itemize}

\subsection{Discussion}\label{Sec:SummAndDiss:Diss}

The confirmation of the existence of a fixed point $\alpha^*$ in the truncated RG flow is very reassuring. Furthermore, the fact that its positions does not appear to change much with variation of the boundary state, fosters hope that this might actually be a genuine feature of the full model, rather than an artefact of the restriction to flat hypercuboid geometries.

We have furthermore investigated the behaviour of the amplitude under vertex translations. This can be understood most easily in terms of whether \emph{large} or \emph{small} spins are preferred by the amplitude. As has already been emphasized in a previous article \cite{Bahr:2015gxa}, the vertex amplitude (without face- and edge-amplitude) scales as $\mathcal{A}_v(j)\sim j^{-21}$, independent of the value of the coupling constants. This is simply a consequence of the Hessian determinant, which governs the behaviour in the large-$j$-asymptotics. Together with the edge amplitudes, which scale as $\mathcal{A}_e\sim j^{3}$, the dressed amplitude (\ref{Eq:DressedAmplitudeStateSum}) of one vertex $v$ scales as
\begin{eqnarray}\label{Eq:Eq:ScalingOfAmplitudes}
\hat{\mathcal{A}}_v^\alpha(j)\;=\;\underbrace{\prod_{f\supset v}\mathcal{A}_{f}^{\frac{1}{4}}}_{\sim j^{12\alpha}}
\underbrace{\prod_{e\supset v}\mathcal{A}_{e}^{\frac{1}{2}}\mathcal{A}_v}_{\sim j^{-9}}.
\end{eqnarray}

\noindent This shows that the behaviour of the amplitude, whether large or small spins are preferred, changes at about $\alpha =\frac{3}{4}$: for smaller $\alpha$, smaller spins are preferred in one vertex, while for larger $\alpha$, larger spins are dominant. A similar dependence on the face amplitude (for three specific values of $\alpha$) was observed for the Barrett Crane spin foam model as well \cite{Baez:2002aw,Khavkine:2007tq}.

In the sum of the RG step, we integrate over a large amount of configurations, all of which have the same $4$-volume\footnote{In this way the RG step is very similar to the prescription in the Causal Dynamical Triangulations approach \cite{Ambjorn:2014gsa}.}. For small $\alpha$, the dominating configurations are such that almost \emph{all} faces have small spins, and only one face carries almost all of the surface area. This is an example of a very irregular subdivision, and is therefore preferred for small $\alpha$. For large $\alpha$ however, very regular subdivisions are preferred, due to the fact that the $4$-volume is fixed, so the state in which the areas are all as large as possible \emph{at the same time} is dominant. This is a heuristic explanation as to why, for small $\alpha$, irregular subdivisions dominate, while for large $\alpha$ regular subdivisions dominate.

Furthermore, it appears that the critical point $\alpha_c$, which separates the ``regular'' and the ``irregular'' phase, lies very close to the fixed point $\alpha^*$. Since at the critical point $\alpha_c$ all subdivisions are, to an extent, weighed equally by the path integral measure, which we have equated with invariance under vertex translation symmetry, this fosters hope that, in the full theory, the two points might indeed be precisely the same. This would mean that the broken diffeomorphism symmetry would be, in fact, restored at the RG fixed point, which has been conjectured for some time \cite{Bahr:2009ku, Bahr:2009qc, Bahr:2011uj}.

It should also be noted that this indicates that the fixed point behaviour observed here would change little under, say, changes of the embedding map (\ref{Eq:EmbeddingMaps}). While our choice (\ref{Eq:EmbeddingMapQuboid}) for this map is only one of many, a moderate change could change the scaling behaviour of the amplitudes (by being absorbed into the edge amplitude, for instance), but this would at most shift the position of $\alpha_c$, and therefore supposedly also $\alpha^*$, but, for instance, not the fact that the two phases exist. \\[5pt]

The prospect of the existence of a non-Gaussian fixed point in the RG flow of spin foam quantum gravity is very exciting, and we feel that this point warrants further investigation. It would be most desirable to make a connection to the asymptotic safety (AS) programme, which postulates the existence of a non-Gaussian fixed point for quantized general relativity.

In the hypercuboidal truncated EPRL model, the RG flow indeed has a fixed point, which shares many features with the one found in AS. In particular, it is UV-attractive. It is also non-Gaussian, in the sense that the fixed point theory is certainly no free theory.

There are several open questions concerning the nature of this fixed point. Most prominently, it would be interesting to establish whether it can be interpreted as being on the boundary between two phases. The model behaves quite differently on both sides of the fixed point, which supports this interpretation, though further analysis is needed to confirm this point. The fact that diffeomorphism symmetry seems to be broken throughout the theory space apart from at the fixed point, hints at a possibility that the continuum symmetries of general relativity could be correctly implemented at this point. This underlines the hope that it could actually be a second order phase transition. We will come back to this point in the future.

Furthermore, it is crucial to extend the analysis by including more intertwiners to the theory, making the truncations less and less severe. That way one could investigate the influence curvature degrees of freedom have on the RG flow. Also, it would allow to make statements about the RG flow of other coupling constants than $\alpha$, such as Newton's constant $\kappa$ and the Barbero-Immirzi parameter $\gamma$.

This way, one could establish whether the properties of the fixed point are merely an artefact of the truncation of the model, or rather a genuine feature of the EPRL model.

\begin{acknowledgments}
This work was funded by the project BA 4966/1-1 of the German Research Foundation (DFG). The authors are indebted to Sebastian Kl\"{o}ser for the introduction to the \verb"Cuba" library.
\end{acknowledgments}

\appendix

\section{The hypercuboidal amplitude}\label{Sec:Appx:Amplitude}

The vertex amplitude $\mathcal{A}_v$ for a hypercuboid depends on its boundary data, i.e. the spins and inertwiners on its boundary graph, depicted in figure \ref{Fig:Hypercuboid_01}. Since the intertwiners are taken to be quantum cuboids (\ref{Eq:CoherentCuboidIntertwiner}), spins of opposing links at a node have to coincide. One can see easily this means that all spins on an ``great circle'' within the boundary graph need to be equal, and since there are six of these, corresponding to the six rotation direction in $4D$, the actual amplitude depends on six spins, corresponding to its six different areas. For an actual four-dimensional hypercuboid, one needs only four numbers to specify all areas, of course. The two excess degrees of freedom are connected to certain non-geometric configurations, which are discussed in detail in \cite{Bahr:2015gxa}.

The amplitude itself, for a value of the Barbero-Immirzi parameter $\gamma < 1$ is factorizing, i.e.~
\begin{eqnarray}
\mathcal{A}_v\;=\;\mathcal{A}_v^+\mathcal{A}_v^-
\end{eqnarray}

\noindent where
\begin{eqnarray}
\mathcal{A}_v^\pm\;=\;\int_{SU(2)^8}d^8g_a\;\prod_{a\to b}\langle -\vec n_{ab}|\,g_a^{-1}g_b\,|n_{ba}\rangle^{(1\pm\gamma)j_{ab}}
\end{eqnarray}

\noindent where the product ranges over links in the boundary graph, which are oriented from cuboid $a$ to cuboid $b$. A vector $\vec n_{xy}$ here is the normal vector in the cuboid $x$, at the square at which it is touching cuboid $y$ (where $x, y=a, b$).

The integral itself is evaluated here in the large $j$-limit. Then, one can use analysis methods detailed in \cite{Barrett:2009gg, Conrady:2008mk}, who have applied these methods to the case of the four-simplex. The generalization to the hypercuboid is straightforward, and it detailed in \cite{Bahr:2015gxa}. The formula for $\mathcal{A}^\pm$ is
\begin{eqnarray}
\mathcal{A}^\pm_v\;=\;\left(\frac{1\pm\gamma}{2}\right)^{\frac{21}{2}}\mathcal{B}_v
\end{eqnarray}

\noindent with
\begin{eqnarray}\label{Eq:TwoStationaryCriticalPoints}
\mathcal{B}_v(j_1,\ldots,j_6)\;=\;\frac{1}{\sqrt{\det H}}\;+\;\text{c.c.}
\end{eqnarray}

\noindent with

\begin{eqnarray*}
\det H\;&=&\;2 \big(j_1^2 (j_2 + j_4) + j_2 j_4 (j_2 + j_4) \\&+&
   j_1 (j_2^2 + (1 + i) j_2 j_4 + j_4^2)\big) \big(j_1^2 (j_3 + j_5) \\&+&
   j_3 j_5 (j_3 + j_5) + j_1 (j_3^2 + (1 + i) j_3 j_5 + j_5^2)\big) \big(j_3 j_4 j_5 \\&+&
   j_2 (j_4 j_5 + j_3 (j_4 + j_5))\big) \big(j_2^2 (j_3 + j_6) + j_3 j_6 (j_3 + j_6) \\&+&
   j_2 (j_3^2 + (1 + i) j_3 j_6 + j_6^2)\big) \big(j_4^2 (j_5 + j_6) \\&+&
   j_5 j_6 (j_5 + j_6) + j_4 (j_5^2 + (1 + i) j_5 j_6 + j_6^2)\big) \big(j_3 j_4 j_6 \\&+&
   j_1 (j_4 j_6 + j_3 (j_4 + j_6))\big) (j_2 j_5 j_6 + j_1 (j_5 j_6 + j_2 (j_5 + j_6)))
\end{eqnarray*}
\noindent It is this asymptotic formula we use for the calculation in our article.

A remark is in order: in \cite{Barrett:2009gg}, the large-$j$-limit relies on the extended stationary phase approximation, for which one needs to compute certain critical, stationary points. In the case of the four-simplex, these exist only if the boundary data $j_{ab}$, $\vec n_{ab}$ satisfy certain conditions. Namely, they have to define a so-called \emph{Regge geometry}, which corresponds to assignments of edge lengths, so that a classical, four-dimensional geometry is defined. In all other cases, the amplitude is exponentially suppressed in the large-$j$-limit.

The situation for the hypercuboid is somewhat different, since, there are solutions for \emph{all} values of $j_1,\ldots j_6$. Up to discrete symmetries, there are two stationary, critical points for $\mathcal{B}_v$, which constitute the two terms in (\ref{Eq:TwoStationaryCriticalPoints}). On these the action is zero, and the amplitude is a pure Hessian determinant. This means that the amplitude is not exponentially suppressed for a large set of non-geometric boundary configurations. They are \emph{dynamically} suppressed, however, when the value of the coupling constant $\alpha$ is above a certain, critical value \cite{Bahr:2015gxa}.

\section{Fadeev-Popov-determinant}\label{Sec:Appx:FadeevPopov}

Due to the fact that for a large region of the coupling constant $\alpha$, one finds that the non-geometric configurations are suppressed \cite{Bahr:2015gxa}, one can assume that the restriction to just geometric states is a decent approximation to the path integral, at least for these values of $\alpha$. Also, there are some arguments why the actual restriction to this subset of states is desirable, since the geometricity conditions appears to be equivalent to the volume simplicity constraint in the hypercuboidal case.

For these reasons, we will perform some of the integrations in this article over the geometric configurations only, i.e.~only over states for which, at each hypercuboid, the six spins $j_1,\ldots,j_6$ satisfy
\begin{eqnarray}\label{Eq:GeometricityConstraintApp}
C_1\;:=\;j_1j_6-j_2j_5\;=\;0\\\nonumber
C_2\;:=\;j_2j_5-j_3j_4\;=\;0.
\end{eqnarray}

\noindent This restriction, however, has to be done carefully. Essentially, the integrals over the $j_f$ in the large $j$-limit have to be constrained to the submanifold $\mathcal{C}$ given by the constraint equations (\ref{Eq:GeometricityConstraintApp}). We do this in such a way that the Riemannian metric on the space of $j_f$, i.e.~$ds^2=\sum_fdj_f^2$, is restricted to $\mathcal{C}$ via pullback.

When the geometricity conditions (\ref{Eq:GeometricityConstraintApp}) are imposed, one can go over from the areas to edge lengths in the path integral (\ref{Eq:SF_StateSum}). Using these, the integral acquires an additional factor, which is the analogue of the Fadeev-Popov-determinant (FPD) in those cases where the constraints in question are actually coming from the presence of gauge symmetries. In what follows we derive the FPD for the case of one $4D$ hypercuboid, and remark that the general case can be easily inferred from the result.

Assume we would like to constrain the integral
\begin{eqnarray}
I\;=\;\int dj_1\ldots dj_6\;F(\vec j)
\end{eqnarray}

\noindent to the subset defined by the constraints (\ref{Eq:GeometricityConstraintApp}), where the integration measure $\mu$ on the constraint hypersurface $\mathcal{C}$ comes from the pull-back of the metric $ds^2=\sum_fdj_f^2$. As coordinates on this surface, the integration variables $j_1,\ldots,j_4$ can be used. One then has
\begin{eqnarray}
j_5\;=\;\frac{j_3j_4}{j_2},\quad j_6\;=\;\frac{j_3j_4}{j_1}.
\end{eqnarray}

\noindent To get the right result when integrating over $j_1,\ldots, j_4$, one needs to incorporate the factor $\frac{1}{\cos\theta}$, where $\theta$ is the angle between the tangent space $T_{\left(j_1,\ldots,j_4,\frac{j_3j_4}{j_2}, \frac{j_3j_4}{j_1}\right)}\mathcal{C}$ and the hyperplane spanned by $\partial_{j_1},\ldots \partial_{j_4}$. In particular, we have

\begin{eqnarray*}
\int_{\mathcal{C}}d\mu\;F\;=\;\int dj_1,\ldots, dj_4\;\frac{1}{\cos\theta}F\left(j_1,\ldots,j_4,\frac{j_3j_4}{j_2}, \frac{j_3j_4}{j_1}\right)
\end{eqnarray*}

\noindent The angle $\theta$ can be computed, and satisfies
\begin{widetext}
\begin{eqnarray}
\cos\theta\;=\;\frac{(j_1^2 j_2^2)}{
\sqrt{
j_1^4 (j_2^2 + j_3^2) (j_2^2 + j_4^2) +
 j_3^2 (j_2^2 + j_3^2) j_4^2 (j_2^2 + j_4^2) +
 j_1^2 (j_3^2 + j_4^2) (j_2^4 + j_3^2 j_4^2)}}.
\end{eqnarray}
\end{widetext}

\noindent In the end, we go over from $j_1,\ldots, j_4$ to length variables $x, y, z, t$, which results in an additional Jacobian
\begin{eqnarray}
J\;=\;\left|\det\frac{\partial(j_1, j_2, j_3, j_4)}{\partial(x, y, z, t)}\right|\;=\;x y^2 z.
\end{eqnarray}

\noindent The final FPD which has to be included in the path integral, when integrating only over geometric configurations, is therefore
\begin{eqnarray}
\Delta_{FP}\;=\;\frac{J}{\cos\theta}.
\end{eqnarray}

\section{Numerical methods}\label{Sec:Appx:Numerical}

In this article, we have used numerical methods to compute the path integral (\ref{Eq:RenormalizedAmplitude}). In a previous article \cite{Bahr:2016hwc} we have relied mostly on simple Markov chain Monte Carlo methods \cite{Diaconis0000aa}, this had severe limitations: For small values of $\alpha$, the integrand develops several ``spikes'', which lie predominantly at the boundary of integration space. This is a consequence of the fact that the path integral prefers highly irregular subdivisions over regular ones. As a result, a random walker would often get ``caught'' in one of these spikes, and not leave it in a numerically reasonable time. This could be explicitly observed, e.g.~in figure 4 of \cite{Bahr:2016hwc}. In particular for small values of $\alpha$, this led to to a high numerical error.

In this article, however, we have made extensive use of state-of-the-art numerical methods, called \verb"Cuba" \cite{Hahn:2004fe}. This comprises a package of several numerical methods, which can be included in various programming languages, such as \verb"C", \verb"Python", \verb"Julia", and even in \verb"Mathematica". The main idea behind these methods (apart from one, which is completely deterministic) are also relying on Monte Carlo methods, while using quite intelligent subdivisions of the integration domain into smaller pieces prior to the integration. This is a major improvement in terms of treating the ``spikes'' in the integrand, and has allowed us to increase the numerical precision of our integrations significantly.

Also, we have made extensive use of Wolfram Alpha's \verb"Mathematica" for basic calculations and the preparation of plots.

\section{Coarse graining hypercuboids: normalization factors}\label{Sec:Appx:Normalization}

Consider a coarse lattice $\Gamma'$. We assume that this lattice is the result of coarse graining a finer lattice $\Gamma$, such that each coarse hypercuboid consists of $2\times 2\times 2\times 2 = 16$ fine hypercuboids. The RG step is designed to produce a vertex amplitude $\hat{\mathcal{A}}_V^{\rm (ren)}$ on $\Gamma'$ from vertex amplitudes $\hat{\mathcal{A}}_v$ on $\Gamma$, such that the expectation values of observables $O(J_F)$ are as close as possible.
\begin{eqnarray}\nonumber
\frac{1}{Z_\Gamma'}\sum_{J_F}O(J_F)\prod_V\hat{\mathcal{A}}^{\rm (ren)}_V\;\approx\;\frac{1}{Z_{\Gamma}}\sum_{j_f}O\left({\scriptstyle\sum_{f\subset F}j_f}\right)\prod_v\hat{\mathcal{A}}_v.\\[5pt]\label{Eq:ExpectationValueCoarseGraining}
\end{eqnarray}

\noindent It will, in general, not be possible to achieve an exact equality here, the reason being that, grouping the fine spins to coarse spins via
\begin{eqnarray}\label{Eq:CoarsegrainingSpins}
J_f\;=\;\sum_{f\subset F}j_f,
\end{eqnarray}
\noindent the rhs of (\ref{Eq:ExpectationValueCoarseGraining}) turns into a nonlocal expression of the coarse spins $J_F$. Explicitly, one has
\begin{eqnarray}\label{Eq:IndexReshuffling}
\sum_{j_f}\prod_v\hat{\mathcal{A}}_v\;&=&\;\sum_{J_F}\sum_{j_f}\delta\left({\scriptstyle\sum_{f\subset F}j_f}\right)\prod_v\hat{\mathcal{A}}_v\\[5pt]\nonumber
&=&\;\sum_{J_F}\left(\sum_{j_f}\delta\left({\scriptstyle\sum_{f\subset F}j_f}\right)\prod_v\hat{\mathcal{A}}_v\right),
\end{eqnarray}

\noindent and the last expression does clearly not factorize over the coarse vertices $V$. This is a general issue, which comes up in many versions of real space renormalization, in particular in tensor networks \cite{levin,guwen}. It is connected to the fact that, in two neighboring coarse hypercuboids $V_1$ and $V_2$, the sum over the fine spins $j_f$ are not independent, if the  face $f$ is a part of both $V_1$ and $V_2$. In other words, the coarse states which are a sum over fine states, contain some entanglement which connects neighboring coarse vertices $V$, so that the resulting expression is not local.

The occurrence of dealing with unwanted entanglement, such as short-range entanglement in tensor network renormalization is an important issue, and is currently discussed in the literature \cite{vidal-evenbly}.

In our case, we remove this entanglement by hand in the RG step, via going over from a sum over \emph{fine states} $j_f$ to a sum over \emph{split fine states} $j_{f,V}$. A split fine state is an assignment of several spins to each fine face $f$, one for every coarse vertex $V$ of which $f$ is a part. We let the sum range over all split fine states $j_{f,V}$, such that
\begin{eqnarray}\label{Eq:ConditionSplitFineStates}
\sum_{f\subset F}j_{f, V}\;=\;J_F\text{ does not depend on $V$.}
\end{eqnarray}

\noindent Then, it is clear that this sum ranges over an expression which factorizes over the $V$, i.e.
\begin{eqnarray}\label{Eq:CoarseGraining01}
\sum_{j_{f,V}}\prod_v \hat{\mathcal{A}}_v\;=\;\sum_{J_F}\prod_V\hat{\mathcal{A}}_V^{\rm (ren)}
\end{eqnarray}

\noindent where in each $\hat{\mathcal{A}}_v$, the $j_f$ are replaced by those $j_{f, V}$ such that $V\supset v$, and
\begin{eqnarray}
\hat{\mathcal{A}}_V^{\rm (ren)}(J_F)\;=\;\sum_{j_{f,V}}\prod_v \hat{\mathcal{A}}_v.
\end{eqnarray}

\noindent Of course, the rhs of (\ref{Eq:CoarseGraining01}) does not equal the original state sum expression on $\Gamma$ given by the lhs of (\ref{Eq:IndexReshuffling}). In an attempt to reduce the error in going over from one to the other, we introduce additional combinatorial factors $C_{J}$, which satisfy
\begin{eqnarray}\label{Eq:SplitFineSum}
\sum_{j_{f,V}}C_J X(J_F)\;=\;\sum_{j_{f}} X(J_F)
\end{eqnarray}

\noindent where $X(J_F)$ is an expression which only depends on $J_F=\sum_{f\subset F}j_f$, and the sum ranges over all split fine states satisfying that (\ref{Eq:ConditionSplitFineStates}) holds. These factors $C_J$ are combinatorial, since the rhs and lhs of (\ref{Eq:SplitFineSum}) contain very different numbers of terms.

To compute these factors, we first note that each coarse face $F$ consists of four fine faces $f$. Given a spin $J$, there are
\begin{eqnarray}
d_J\;:=\;\left(\begin{array}{c}2J+3\\3\end{array}\right)
\end{eqnarray}

\noindent possibilities for half-integer fine spins $j_1,\ldots, j_4$ to add up to $J$. If we define $N_A$ to be the number of possible fine states given a certain coarse state, then we have
\begin{eqnarray}
N_A\;=\;\prod_{\text{independent }F}d_{J_F}.
\end{eqnarray}

\noindent Here the sum ranges over \emph{independent} coarse faces, i.e.~all faces which are forced to have equal spin by the hypercuboidal symmetry, are counted as one face. These are all faces that arise by translations on the lattice orthogonal to that face.

Furthermore, let $N_B$ be the number of split fine states for a given coarse state, then, on the hypercuboidal lattice, we have that
\begin{eqnarray}
N_B\;=\;\prod_V\big(\prod_{i=1}^6d_{J_i}\big),
\end{eqnarray}

\noindent where $i$ ranges over the six independent faces in the hypercuboid dual to the vertex $V$.

Now assume that the lattice $\Gamma'$ consists of of $N_1\times N_2\times N_3\times N_4$ hypercuboids. In that case, there are
\begin{eqnarray*}
M\;:=\;N_1N_2+N_1N_3+N_1N_4+N_2N_3+N_2N_4+N_3N_4
\end{eqnarray*}

\noindent independent coarse faces, while there are $R:= N_1N_2N_3N_4$ coarse vertices. So, the rhs of (\ref{Eq:CoarseGraining01}) overcounts the states (in the case of all $J_F\equiv J$ equal) by a factor of
\begin{eqnarray}
\frac{N_B}{N_A}\;=\;(d_J)^{6R-M}
\end{eqnarray}

\noindent with respect to the lhs of (\ref{Eq:IndexReshuffling}). So for each vertex the sum over split fine states should be scaled down by $(d_J)^{6-M/R}$. For large lattices, the number $R$ grows much faster than $M$, which is why we approximate $M/R\approx 0$, and write
\begin{eqnarray}\nonumber
\frac{1}{Z_\Gamma'}\sum_{J_F}O(J_F)\prod_V\hat{\mathcal{A}}^{\rm (ren)}_V\;\approx\;\frac{1}{Z_{\Gamma}}\sum_{j_f}O\left({\scriptstyle\sum_{f\subset F}j_f}\right)\prod_v\hat{\mathcal{A}}_v\\[5pt]\label{Eq:FinalFormulaApprox}
\end{eqnarray}

\noindent with
\begin{eqnarray}
\hat{\mathcal{A}}_V^{\rm (ren)}(J_F)\;=\;\frac{1}{\prod_{F=1}^6d_F}\sum_{j_{f,V}}\prod_v \hat{\mathcal{A}}_v.
\end{eqnarray}

\noindent The approximation (\ref{Eq:FinalFormulaApprox}) is in finite-size effects of the lattice, as well as the suppression of the non-locality. In other words, for amplitudes $\hat{\mathcal{A}}_v$ which only depend on $J_F$ via (\ref{Eq:CoarsegrainingSpins}), and not on $j_f$, the approximation becomes exact in the limit of large lattices.

\bibliography{bibliography}

\end{document}